# Hamiltonian structure of three-dimensional gravity in Vielbein formalism

Mahdi Hajihashemi[1,*] and Ahmad Shirzad[1,2,†]

[1]*Department of Physics, Isfahan University of Technology, P.O.Box 84156-83111 Isfahan, Iran*
[2]*School of Particles and Accelerators, Institute for Research in Fundamental Sciences (IPM), P.O.Box 19395-5531 Tehran, Iran*

Considering Chern-Simons like gravity theories in three dimensions as first order systems, we analyze the Hamiltonian structure of three theories Topological massive gravity, New massive gravity, and Zwei-Dreibein Gravity. We show that these systems demonstrate a new feature of the constrained systems in which a new kind of constraints emerge due to factorization of determinant of the matrix of Poisson brackets of constraints. We find the desired number of degrees of freedom as well as the generating functional of local Lorentz transformations and diffeomorphism through canonical structure of the system. We also compare the Hamiltonian structure of linearized version of the considered models with the original ones.

## I. INTRODUCTION

It is well known that, in three dimensions, the Einstein-Hilbert action of general relativity has no dynamical degree of freedom. However, this is not the case for other extensions of the theory [1,2]. Deser, Jakieve, and Tempelton in a pioneer paper showed that adding a special Chern-Simons term with respect to Christoffel symbols leads to a true dynamical theory in three dimensions [3]. This theory, known as topological massive gravity (TMG), is the basis of some other suggested models in three dimensions. In this paper we want to study the dynamics of these models in the framework of Hamiltonian formalism. Going to the vielbein framework, one is able to introduce the scalar curvature and the cosmological constant term of the of Hilbert-Einstein theory, as well as the added Chern-Simons term, in terms of the corresponding differential forms. In Sec. III we will review the general structure of these models recognized as the Chern-Simons-like theories (CSL).

The dynamical content of the CSL theories in general, and the TMG in particular, has been the subject of a number of papers [4–9]. In particular, the number of dynamical degrees of freedom is not so obvious. In the original paper, Deser, Jackiw, and Templeton linearized the action of TMG around the flat solution and showed that TMG contains one propagating scalar massive mode [3]. This strong observation brings some authors to the conclusion that the TMG in its original (i.e., non linearized) version possesses one dynamical degree of freedom, as well.

It is more or less agreed on that the most accurate method to find the dynamical content of a theory, including the number of degrees of freedom, is the Hamiltonian (i.e., canonical) analysis. Hence, a great deal of discussions about the dynamical structure of TMG, as well as other CSL theories, are given in the framework of Hamiltonian formulation. However, due to complications of the algebra of constraints, a satisfactory investigation about canonical structure and dynamical behavior of the variables may be lost or hidden behind complicated calculations.

Some of the exciting works do not agree with each other. For example in [5] the expression of secondary constraints is different from [7]; and the number of constraints as well as the number of degrees of freedom for ZDG is different in [10,11].

Some of the works are done in the context of ordinary Dirac formalism in which the dynamical variables such as vielbein $e_\mu^a$, spin connection $\omega_\mu^a$, etc. are considered as configuration space variables. In this method, we have a duplicated phase space with so many coordinates and so many constraints. For example, for TMG one encounters a phase space with 54 variables and 40 constraints in four levels [6]; while in "first order formalism" we have a phase space of 18 variables and ten constraints in just one level. Besides difficulties resulted from introducing so many unnecessary variables, this method has not been applied to all CSL theories. It is not also showed to be equivalent to first order formalism.

It is more or less familiar to authors of this field that in other to find the correct number of degrees of freedom, as well as the correct dynamical behavior (i.e., equivalent to direct Euler-Lagrange equations of motion), one needs to introduce some additional constraints. They try to find these constraints, through ordinary Dirac methods, as secondary constraints [8,12]. However, as we will show in details in Secs. II and III, the CSL theories exhibit a different feature of constrained systems, in which a dynamical system may have different properties in a specific subregion of phase space. This may include additional gauge symmetries which we call "limited gauge transformations." This subregion should be specified be definite factors of the determinant of the matrix of Poisson brackets of second class constraints.

[*]mehdi.hajihashemi@ph.iut.ac.ir
[†]shirzad@ipm.ir

We think that this new aspect of constrained system, which is explained in Sec. II, is less understood in the literature. However, the "new constraints" emerged by this procedure are different from the ordinary secondary constraints, and the Hamiltonian structure differs from systems in which the matrix of Poisson brackets of constraints has constant rank throughout the whole phase space.

Most of the works done end up with determination of the number of degrees of freedom [6,8,12]. However, the canonical analysis may have more products. It is also well known that the gauge symmetries of a theory may be better understood in the algebraic structure of first class constraints of the system [13–15]. In this regard, there are not so many results in the literature on three dimensional gravity. Hence, another aim of this paper is to give a satisfactory discussion about the symmetries of the CSL theories in the framework of Hamiltonian analysis.

As is well know, every gravitational theory is symmetric under arbitrary change of the space-time coordinates (i.e., diffeomorphism). On the other hand, in the framework of vielbein formalism, at the expense of increasing the number of variables, we add the symmetry under local Lorentz transformation (LLT). In three dimensions each of these gauge transformations introduce 3 arbitrary infinitesimal gauge parameters. In the Hamiltonian structure of the system, one needs three first class constraints (in each generation of the constraints) to act as generators of the gauge transformations in each case. This task is not done completely and in all of the CSL theories yet. Specially the first class-ness of the corresponding constraints is not clarified explicitly.

Our final aim is to compare the Hamiltonian structure of the linearized version of CSL theories around the flat metric with the original theories. Since the fluctuation is done within the physical sector of the theory, we find that the constraint structure of the linearized models are similar to the original one with no need to introducing new constraints.

In Sec. II we give a discussion about first order systems and the less-notified possibility of emerging new constraints which are different from ordinary secondary constraints. In Sec. III we review the CSL theories and give the common features of these models. In Secs. IV–VI we give a detailed discussion about the Hamiltonian structures of TMG, new massive gravity (NMG) and zwei-dreibein (ZDG) gravity respectively. It worth emphasizing that ZDG, in its own right, is very attractive and its Hamiltonian structure had been the subject of some important papers [11,12].

In Appendix A we give the results of heavy calculations of the Poisson brackets. In Appendix B we give the gauge transformations of the dynamical variables of the considered models by using the gauge generators derived in the text. It should be noted that for similar CSL theories such as MMG [16] and GMG [17], the essential features are very similar to the models considered here and no new result arises.

## II. FIRST ORDER SYSTEMS

In this section we show the general structure of a first order Lagrangian. We present the formalism for a system with finite number of degrees of freedom; generalization to a field theory is easily understood. Assume a dynamical system described by the set of dynamical variables $y^i(t)$, $i = 1, 2, \ldots N$ and $u_\alpha(t)$ $\alpha = 1, 2, \ldots m$, where the dynamics of the theory is given by the following first order Lagrangian

$$L = a_i(y)\dot{y}^i - u^\alpha \phi_\alpha(y) - H(y). \quad (1)$$

The Euler-Lagrange equations of motion read

$$\phi_\alpha(y) = 0, \quad (2)$$

$$\omega_{ij}\dot{y}^i = u^\alpha \frac{\partial \phi_\alpha}{\partial y^j} + \frac{\partial H}{\partial y^j}, \quad (3)$$

where

$$\omega_{ij} \equiv \frac{\partial a_i}{\partial y^j} - \frac{\partial a_j}{\partial y^i} \quad (4)$$

is the symplectic matrix. Assuming $\omega$ is nonsingular (for singular $\omega$ see Ref. [18]), we can construct the Poisson brackets among the canonical variables $y^i$ as

$$\{y^i, y^j\} = \omega^{ij}, \quad (5)$$

where $\omega^{ij}$ is the inverse of $\omega_{ij}$. In this way the equations of motion (3) can be considered as

$$\dot{y}^i = \{y^i, H_T\}, \quad (6)$$

where

$$H_T = u_\alpha \phi_\alpha + H. \quad (7)$$

Hence, $u_\alpha$ can be viewed as Lagrange multipliers corresponding to the primary constraints $\phi_\alpha$ and $H$ can be considered as the canonical Hamiltonian. The equations of motion (6) can be generalized for every function $f(y)$ as

$$\dot{f} = \{f, H_T\}. \qquad (8)$$

Specially we can investigate the consistency of the constraints $\phi_\alpha(y)$ as follows

$$\dot{\phi}_\alpha = 0 \Rightarrow \{\phi_\alpha, \phi_\beta\}u_\beta + \{\phi_\alpha, H\} \approx 0, \qquad (9)$$

where the symbols $\approx$ means weak equality, i.e., equality on the constraint surface $\phi_\alpha \approx 0$. Equations (9) can be viewed as equations for determining the Lagrange multipliers $u_\alpha$. If the matrix $M^0_{\alpha\beta} = \{\phi_\alpha, \phi_\beta\}$ is full rank, all of the Lagrange multipliers would be determined in terms of the canonical variables. However, if $m^0 \equiv \text{rank} M^0 < m$, there are $m - m^0$ undetermined Lagrange multipliers which correspond to combinations $\phi_A$ of $\phi_\alpha$'s such that

$$\{\phi_A, \phi_\alpha\} \approx 0 \quad A = 1, \ldots, m - m_0 \quad \alpha = 1, \cdots m. \qquad (10)$$

The constraints $\phi_A$ should be recognized as first class constraints which are generators of the gauge transformations. However, for the constraints $\phi_A$, the second terms in (9), i.e., $\{\phi_A, H\}$ give rise to secondary constraints. Then one should go through consistency of secondary constraints. This procedure may lead to determining some Lagrange multipliers, which convert the corresponding secondary constraints and their associated primary constraints to second class constraints (find the details in Ref. [18]).

Fortunately, in this paper we do not need to consider such details. Our chance in CSL theories is that $H = 0$, as can be seen from the general form of the Lagrangian (35) below. Henceforth we assume this is the case. So consistency equations (9) read

$$\{\phi_\alpha, \phi_\beta\}u_\beta = 0. \qquad (11)$$

As is seen, the consistency of first class constraints $\phi_A$ is satisfied identically and do not lead to new constraints, i.e., for these theories we have not secondary constraints at all. At this stage the existing first class constraints are generators of gauge transformations through the whole phase space. We call this kind of gauge symmetries *unlimited gauge transformations* in contrast to *limited gauge transformations* to be discussed below. The number of parameters of the unlimited gauge transformations is

$$G_u = m - m_0, \qquad (12)$$

which is the same as the number of first class constraints $\phi_A$, as well as the number of $u_A$'s remained undetermined.

The remaining constraints $\phi_a$, $a = 1, \cdots m_0$, are (so far) second class in the sense that the matrix $M^1_{ab} \equiv \{\phi_a, \phi_b\}$ of their Poisson brackets is non singular. So the nontrivial part of the consistency equations (11) is given as

$$M^1_{ab}u_b = 0. \qquad (13)$$

Equation (13) is the most important requirement of the consistency of the constraints. The simplest solution of this equation is $u_a = 0$ which is valid throughout the whole phase space. This choice is not acceptable in the CSL theories as we will see later. However, we can find other possibilities by considering the determinant of the matrix $M^1$. Singular properties of the matrix $M^1$ in definite subregions of the phase can be achieved by considering the solutions of the equation

$$\det M^1 \equiv f(y) = 0. \qquad (14)$$

Suppose the function $f(y)$ is a polynomial in variables $y^1$ to $y^N$ and it may be factorized as follows

$$f(y) = \prod_k h_k(y) \prod_c g_c(y), \qquad (15)$$

where the polynomials $g_c(y)$ are supposed to be nonvanishing throughout the whole phase space (like $(y^1)^2 + (y^2)^2 + \cdots + a^2$) while each polynomial $h_k(y)$ defines a nontrivial surface in the phase space as $h_k(y) = 0$. We denote the factors $h_k$ as *vanishable* factors and the $g_c$ as *nonvanishable* factors of the function $f(y)$. Remember that we do not consider complex solutions in the context of gravity theories.

The factorized polynomials $h_k(y)$ may occur in Eq. (15) more than once. The rank of $M^1$ is decreased on a subregion of phase space given by a union of some subset of $h_k(y)$'s. For example if $f(y)$ contains factors $[h_1(y)]^3[h_2(y)]^2$ in the right-hand side of Eq. (15), rank$M^1$ is decreased by 3 on the surface $h_1(y) = 0$, by 2 on the surface $h_2(y) = 0$ and by 5 on the union of the two surfaces.

Suppose we have chosen to land on the limited subregion of the phase space given by the relations

$$h_r = 0 \quad r = 1, \cdots K, \qquad (16)$$

which may or may not cover the whole set of vanishable factors $h_k$. Suppose $\tilde{M} \equiv M_{h_r = 0}$ and rank $\tilde{M} = \tilde{m} < m_0$. Then from Eq. (13) we have $\tilde{m}$ independent equations for $m_0$ Lagrange multipliers $u_a$ as unknowns. Hence, $m_0 - \tilde{m}$ of the Lagrange multipliers turn out to be temporarily undetermined on the surface $h_r = 0$ and the same number of second class constraints have been temporarily converted to first class.

However, the new constraints $h_r$ should not be violated by the dynamics of the system. So we should demand

$\{h_r, H_T\} = 0$, where $H_T = u_\alpha \phi_\alpha$. Our claim, to be proved later, is $\{h_r, \phi_A\} \approx 0$. Hence, consistency of $h_r$ implies

$$\{h_r, \phi_a\} u_a \approx 0 \quad r = 1, \ldots, K, \quad a = 1, \cdots m_0. \quad (17)$$

Equation (17) leads to determination of some of the $u_a$'s. One may consider the possibility that consistency of some of $h_r$'s is satisfied identically on the constraint surface. However, in the case of the Chern-Simons like theories, considered in this paper, this is not the case and exactly $K$ combinations of $u_a$'s would be determined from Eq. (17). This means that the new constraints $h_r$ constitute a system of second class constraints with $K$ of the constraints $\phi_a$ which have been temporarily converted to first class constraints on the surface $h_r = 0$. Hence, $K$ of the converted first class constraints convert back to be second class constraints.

Altogether, the algebra of the constraints have been changed on the surface $h_r = 0$; so that some of the second class constraints are converted finally to first class ones and have not converted back (in the process of consistency of new constraints $h_r$) to second class. This kind of first class constraints, which are $m_0 - \tilde{m} - K$ in number, can be considered as the generators of a new kind of gauge symmetry, valid only on the limited subregion $h_r = 0$. We call this kind of gauge symmetry "limited gauge transformations." So the number of undetermined Lagrange multipliers at this stage is

$$G_l = m_0 - \tilde{m} - K, \quad (18)$$

where $G_l$ denotes the number of parameters of limited gauge transformations.

The second class constraints, on the other hand, are composed of $\tilde{m}$ constraints remained in the reduced matrix $\tilde{M}$, $K$ converted back second class constraints and finally the $K$ new constraints $h_r$. Hence, the total number of final second class constraints reads

$$S = 2K + \tilde{m}. \quad (19)$$

To check the results, the total number of constraints is

$$G_u + G_l + S = m + K, \quad (20)$$

which is the total number of original and new constraints. The well-known formula for the number of dynamical degrees of freedom of a constrained system is [19]

$$D = N - 2F - S, \quad (21)$$

where $N$ is the dimension of the original phase space, $F$ is the number of first class constraints and $S$ is the number of second class constraints. For the systems under consideration in this section $F = G_u + G_l$ and the total number of dynamical degrees of freedom reads

$$D = N - 2(G_u + G_l) - S = N - 2m + \tilde{m}. \quad (22)$$

As we will show later the constraints structure of CSL theories coincide exactly with what we said here.

Let us give a proof for our claim given above about Poisson brackets of the new constraints $h_r(y)$ with the existing constraints. Since the number of second class constraints is even, the matrix $M^1$ is even dimensional. For simplicity assume $\det M^1 = f(y)$ has only one vanishable factor $h(y)$, i.e., suppose $K = 1$. This may happen only if one row (say the first row) of $M^1$ is proportional to $h(y)$. However, since $M^1$ is antisymmetric, the first column should also be proportional to $h(y)$.

Now suppose the matrix $\bar{M}^1$ is the $(m_0 - 1) \times (m_0 - 1)$ matrix derived by omitting the first row and first column of $M^1$. This is an antisymmetric matrix of odd dimension, which cannot be non-singular. Hence necessarily we should have $\det \bar{M}^1|_{h=0} = 0$. This means that in principle one may redefine the constraints such that another row and column again be proportional to $h(y)$. So at one hand, the power of $h(y)$ in $\det M^1$ should be at least two, and on the other hand, the rank of $M^1$ would be decreased at least by two. In other words, $\tilde{m} \equiv \text{rank} M^1$ depends on the power of $h(y)$ in $\det M^1$. If the power is 4 then $\tilde{m} = m^0 - 4$, etc. In this way one deduces that the rank of $M^1$ should always decrease by an even number on the surface $h_r(y) = 0$ of the new constraints. Remembering that $m^0$, which is the number of the original second class constrains, is even, we are happy to find that $\tilde{m}$ is always even.

Suppose then we have redefined the constraints $\varphi_a$ such that for the first two constraints $\varphi_1$ and $\varphi_2$ we have

$$\{\varphi_1, \varphi_a\} = h\chi_a, \quad (23)$$

$$\{\varphi_2, \varphi_a\} = h\psi_a. \quad (24)$$

Let us first investigate vanishing of $\{\phi_A, h(y)\}$. From Jacobi identity we have

$$\{\phi_A, M^1_{12}\} = \{\varphi_1, \{\phi_A, \varphi_2\}\} - \{\varphi_2, \{\phi_A, \varphi_1\}\}. \quad (25)$$

In the first term on the right-hand side of Eq. (25), the Poisson bracket $\{\varphi_A, \varphi_2\}$ is either first class which would have vanishing Poisson bracket with $\varphi_1$, or is second class which would have vanishing Poisson bracket with $\varphi_1$ on the surface $h(y) = 0$ according to Eq. (23). This implies

$$0 \approx \{\phi_A, M^1_{12}\} = \{\phi_A, h\chi_2\}, \quad (26)$$

which is weakly equal to $\chi_a\{\phi_A, h\}$ on the surface $h = 0$. This proof establishes our claim that $\{\phi_A, h(y)\} \approx 0$.

Now let us investigate the possibility that $\{h(y), \varphi_b\}$ may not vanish for some $\varphi_b$. On the surface $h(y) = 0$ we can consider

$$\{\{\varphi_1, \varphi_a\}, \varphi_b\} \approx \chi_a \{h, \varphi_b\}. \tag{27}$$

Since $\varphi_1$ is second class $\{\varphi_1, \varphi_a\}$ should not vanish weakly for all $\varphi_a$. So, in principle, it is possible that $\{\varphi_1, \varphi_a\}$ have nonvanishing Poisson bracket with some $\varphi_b$. Hence, the new constraint $h$, as well, may have nonvanishing Poisson bracket with some $\varphi_b$. However, consistency of $h$ (or each $h_r$) can at most convert one first class constraint back into second class. In other words, if the Poisson bracket of $h$ with more than one constraint is nonzero, it is possible to redefine the constraints in such a way that only one of them is conjugate to $h$.

For more details, consider a system of second class constraints $\chi, \Pi_1, \Pi_2, \ldots$ where $\{\chi, \Pi_i\} = C_i$ and $C_i$ depends on non constraint variables. Then one can choose $\Pi_1$ say, as the conjugate constraint to $\chi$ and redefine the other constraints $\Pi_i$ as $\tilde{\Pi}_i = \Pi_i - \frac{C_i}{C_1} \Pi_1 \approx \Pi_i$. Then $\{\chi, \tilde{\Pi}_i\} \approx 0$. For example in the system of $\varphi_1 = q_1, \varphi_2 = p_1, \varphi_3 = p_1 + q_2$ and $\varphi_4 = p_1 + p_2$ the constraints $\varphi_2, \varphi_3$ and $\varphi_4$ are seemingly conjugate to $\varphi_1$, but one can choose just $\varphi_2$ as the conjugate constraint to $\varphi_1$ and redefine the other constraints as $\tilde{\varphi}_3 = \varphi_3 - \varphi_2 = q_2$ and $\tilde{\varphi}_4 = \varphi_4 - \varphi_2 = p_2$. Clearly $\tilde{\varphi}_3$ and $\tilde{\varphi}_4$ are no longer conjugate to $\varphi_1$. This shows that it is always possible to remove the effect of the conjugate constraint (of the second class constraint $\chi$) from all other constraints, such that $\chi$ is conjugate only to its own partner.

Now consider the following Jacobi identity

$$\{\varphi_1, \{\varphi_2, \varphi_a\}\} + \{\varphi_2, \{\varphi_a, \varphi_1\}\} + \{\varphi_a, \{\varphi_1, \varphi_2\}\} = 0, \tag{28}$$

Using Eqs. (23) and (24) this equation, on the surface $h(y) = 0$, read

$$\{\varphi_1, h\}\psi_a - \{\varphi_2, h\}\chi_a + \{\varphi_a, h\}\chi_2 \approx 0. \tag{29}$$

According to our discussion in the last paragraph, the last term of Eq. (29) does not vanish at least for one $\varphi_a$. Hence, there exist at least one index $a$ such that

$$\{\varphi_1 \psi_a - \varphi_2 \chi_a, h\} \approx \{\varphi_1, h\}\psi_a - \{\varphi_2, h\}\chi_a \neq 0. \tag{30}$$

This means that the new constraint has non vanishing Poisson bracket with some combination of the constraints $\varphi_1$ and $\varphi_2$ which were second class constraints converted to first class on the surface of new constraint $h(y)$.

### III. CHERN-SIMONS-LIKE THEORIES

The Chern-Simons-like theories are a set of models which are extension of 3D gravity, in the vielbein formalism [20]. In this framework, we use the orthogonal non-coordinate basis $\hat{e}_a \equiv e^\mu_a \partial_\mu$ for tangent space at an arbitrary point, such that $\hat{e}_a \cdot \hat{e}_b \equiv g_{\mu\nu} e^\mu_a e^\nu_b = \eta_{ab}$. The corresponding basis for cotangent space are $\hat{\theta}^a \equiv e^a_\mu dx^\mu$ with the following property

$$g_{\mu\nu} = e^a_\mu e^b_\nu \eta_{ab}. \tag{31}$$

We refer to $\mu, \nu, \ldots$ as the curved space indices and $a, b, \ldots$ as the flat space indices. The next set of dynamical variables are spin connections, which in arbitrary dimensions are two-forms. Since in three dimensions every two form is Hodge dual of a one-forms, one defines the covariant derivative of an arbitrary vector as

$$D_\mu V^a = \partial_\mu V^a + \varepsilon^a_{bc} \omega^b_\mu V^c, \tag{32}$$

where three one-forms $\omega^b$ are (Hodge dual of) spin connections. The torsion and curvature tensors in this formalism are the following two forms

$$T(\omega) = de + \omega \times e, \qquad R(\omega) = d\omega + \frac{1}{2}\omega \times \omega, \tag{33}$$

where the cross symbol means contraction with the use of $\varepsilon^a_{bc}$ (of the flat space indices), such that for instance the first equation of (33) read

$$T^a_{\mu\nu} = \partial_{[\mu} e^a_{\nu]} + \varepsilon^a_{bc} \omega^b_{[\mu} e^c_{\nu]}. \tag{34}$$

In CSL theories we need also to consider some auxiliary fields as dynamical variables. For example in TMG and MMG we need to impose the torsion free condition by hand, i.e., by adding the 3-form term $\eta_{ab} h^a T^b$ to the Lagrangian. Hence, the variables $h^a_\mu$ are also invited. It is customary in the literature to call different one-form field variables, such as $e^a$, $\omega^a$ and $h^a$ as different flavors of the theory and enumerate them as $a^{ra}_\mu$ where r, s, t, ... are flavor indices. Assuming $F$ to be the number of flavors, a CSL theory contains $3 \times 3 \times F = 9F$ variables. Following the notations of Ref. [12], the Lagrangian as a three form in this formalism may be written as

$$L = \frac{1}{2} g_{rs} a^r \cdot da^s + \frac{1}{6} f_{rst} a^r \cdot (a^s \times a^t), \tag{35}$$

where wedge product between forms is implicit and dot means contraction of upper flat space indices with $\eta_{ab}$. Each CSL theory is characterized by symmetric coefficients $g_{rs}$ and $f_{rst}$ with no geometrical or algebraic meaning. The equations of motion resulting from the Lagrangian (35) read

$$g_{rs} da^{sa} + \frac{1}{2} f_{rst} (a^s \times a^t)^a = 0. \tag{36}$$

Using the definition $a^{ra} = a^{ra}_\mu dx^\mu$ we can separate time and space indices of the curved indices as follows

$$a^{ra} = a_0^{ra} dt + a_i^{ra} dx^i. \quad (37)$$

Hence, the first order Lagrangian density of the CLS theories read

$$\mathcal{L} = -\frac{1}{2} \varepsilon^{ij} g_{rs} a_i^r \cdot \dot{a}_j^s + a_0^r \cdot \phi_r, \quad (38)$$

where $\varepsilon^{ij} \equiv \varepsilon^{0ij}$ is the spatial part of Levi-Civita symbol and the set of $3F$ functions

$$\phi_r^a = \varepsilon^{ij} \left( g_{rs} \partial_i a_j^{sa} + \frac{1}{2} f_{rst}(a_i^s \times a_j^t)^a \right) \quad (39)$$

are considered as primary constraints in the first order formalism. As is apparent, $3F$ variables $a_0^{ra}$ should be considered as Lagrange multipliers, while the phase space variables are constituted from $a_i^{ra}$ which are $6F$ in number. From the first term of Eq. (38) we can read the symplectic matrix as

$$\Omega_{ar,bs}^{ij} = g_{rs} \eta_{ab} \varepsilon^{ij}. \quad (40)$$

The inverse of the symplectic matrix gives the Poisson brackets of our dynamical variables [see Eq. (5)] as

$$\{a(\mathbf{x})_i^{ra}, a(\acute{\mathbf{x}})_j^{sb}\} = g^{rs} \eta^{ab} \varepsilon_{ij} \delta^2(\mathbf{x} - \acute{\mathbf{x}}), \quad (41)$$

where $g^{rs}$ is the inverse of $g_{rs}$. Since there is no canonical Hamiltonian in the left-hand side of Eq. (38), the (density of) total Hamiltonian is just a linear combination of the primary constraints $\phi_r^a$ as follows

$$\mathcal{H}_T = -a_0^r \cdot \phi_r. \quad (42)$$

In this way, the consistency of primary constraints will not give any secondary constraint.

To investigate consistency of the constraints (39), the Poisson brackets of the constraints with the total Hamiltonian (42) gives

$$\int d^2 \mathbf{x}' \{\phi_r^a(\mathbf{x}), \phi_s^b(\mathbf{x}')\} a_0^{sb}(\mathbf{x}') \approx 0. \quad (43)$$

The matrix of Poisson brackets of constraints, i.e., the matrix $M$ of Sec. II, is the following $3F \times 3F$ matrix

$$M_{ra,sb} = \{\phi_r^a, \phi_s^b\}. \quad (44)$$

Calculating the elements of $M_{ra,sb}$ is a cumbersome process. Since the constraints (39) contain spatial derivatives of the variables, one should care about the derivatives of delta functions. Some authors use the smearing method to avoid difficulties. However, we prefer to do an explicit calculation to see directly the real algebra of Poisson brackets among the constraints. In Appendix A we give the technical details about evaluating the Poisson brackets of different terms of the constraints.

Depending on the rank of the matrix $M$ in Eq. (43) some of Lagrange multipliers may remain undetermined. As we discussed generally in the previous section, these correspond to first class constraints which are generators of unlimited gauge transformations. We will see soon that for CSL theories this symmetry is the LLT. In TMG and NMG, with no need to any new condition, the constraints $\phi_\omega$ are first class and generate LLT. In ZDG, which we have two sets of vierbeins and spin connections, the combination $\phi_W = \phi_\omega + \phi_{\omega'}$ have the same role.

Then we should investigate the non singular part of Eq. (43) as in the general case of Eq. (13). Since $\det M \neq 0$, Eq. (43) leads to $a_0^{sb}(\mathbf{x}') = 0$, for some flavors, on the whole phase space, except regions where $\det M$ vanishes. This may be unacceptable physically, since $e_0^a = 0$, for instance, leads to a singular metric. To avoid this nonsense result, one needs to investigate the singularity properties of the matrix $M$ in special subregions of the phase space. This makes us to consider new constraints of the kind $h_r$, mentioned in the last section. This should be done case by case in different models. As we will see, in physical subregions of the phase space the diffeomorphism symmetry arise as limited gauge transformations.

Let us see what is the number of dynamical variables for CSL theories. As we said before, in CSL theories the number of phase space variables is $N = 6F$ and the number of primary constraints is $m = 3F$. Now remembering the general result of Eq. (22) we have

$$D = 6F - 2 \times 3F + \tilde{m} = \tilde{m}. \quad (45)$$

It means that in the CSL theories the number of degrees of freedom is rank of matrix $\tilde{M} \equiv M|_{h_r=0}$.

For future use let us quote a formula from Ref. [12] which can be found by taking the exterior derivative of the equations of motion and then using the same equations, i.e.,

$$f_{\mathrm{q[r}}^{\mathrm{t}} f_{\mathrm{s]pt}} a^{ra} a^{\mathrm{p}} \cdot a^{\mathrm{q}} = 0. \quad (46)$$

This equation is consistent with our canonical approach and can be shown to be equivalent to Eq. (43) above. Similar equation are also given in Ref. [21]. However, for a fixed index s, another equation is deduced from Eq. (46) in this references as

$$f_{\mathrm{q[r}}^{\mathrm{t}} f_{\mathrm{s]pt}} a^{\mathrm{p}} \cdot a^{\mathrm{q}} = 0. \quad (47)$$

This equation is then considered as secondary constraints of the CSL theory. Two points should be noticed here. First, secondary constraint has a special meaning in the literature of constrained system, which implies definite properties. A secondary constraint is derived as the Poisson bracket of a primary constraint with the canonical Hamiltonian. This is not characteristics of the Eq. (47). On the other hand it is not legitimate to consider the constraints (47) as preassumed constraints, since this can not be done covariantly by adding a term to the Lagrangian.

Second, Eq. (47) can not be deduced automatically from Eq. (46); it needs to definite assumptions about invertibility of the variables. However, it needs care that, invertibility of vielbein does not imply by itself singularity of the matrix $M$. It can be correct on the other way, i.e., if there exist a possibility for $M$ to be singular, then there is a possibility for the vielbeins to be nonvanishing. In fact the constraints derived in the form of Eq. (47), which are equivalent to the new constraints $h_r$, are originated from the singular properties of the matrix $M$ on special subregions of the phase space, as discussed in details in this paper.

## IV. HAMILTONIAN STRUCTURE OF TMG

The theory of topological massive gravity is achieved by adding a Chern-Simons term to the Lagrangian of the ordinary GR [3]. The canonical analysis of TMG is the subject of several papers [22–25]. In the framework of vielbein formalism the Lagrangian of TMG reads [12]

$$L = -\sigma e.R(\omega) + h.T(\omega) + \frac{1}{2\mu}\left(\omega.d\omega + \frac{1}{3}\omega.\omega \times \omega\right), \quad (48)$$

where $\sigma$ is a sign and $\mu$ is a mass parameter of the model. In the Lagrangian (48) the auxiliary fields $h_\mu^a$'s are Lagrange multipliers introduced to enforce the torsion free condition. Hence, vielbein $e$, spin-connection $\omega$ and the auxiliary fields $h$ are considered as three flavors. The nonvanishing coefficient $g_{rs}$ and $f_{rst}$ peculiar to this model can be read from the Lagrangian (48) as

$$g_{e\omega} = -\sigma \quad g_{eh} = 1 \quad g_{\omega\omega} = \frac{1}{\mu}$$

$$f_{e\omega\omega} = -\sigma \quad f_{eh\omega} = 1 \quad f_{\omega\omega\omega} = \frac{1}{\mu}. \quad (49)$$

Before investigating the canonical structure of the model, let us write directly the Euler-Lagrange equations of motion resulted from the Lagrangian (48), i.e.,

$$T(\omega) = 0,$$
$$\sigma R(\omega) - dh - \omega \times h = 0,$$
$$R(\omega) - \mu e \times h + \sigma\mu T(\omega) = 0. \quad (50)$$

Following the recipe given in Eqs. (38) and (39), the Lagrangian density reads

$$\mathcal{L} = \frac{\sigma}{2}\epsilon^{ij}e_i.\dot{\omega}_j + \frac{\sigma}{2}\epsilon^{ij}\omega_i.\dot{e}_j - \frac{1}{2}\epsilon^{ij}h_i.\dot{e}_j - \frac{1}{2}\epsilon^{ij}e_i.\dot{h}_j$$
$$- \frac{1}{2\mu}\epsilon^{ij}\omega_i.\dot{\omega}_j + e_0.\phi_e + \omega_0.\phi_\omega + h_0.\phi_h, \quad (51)$$

where the primary constraints are as follows

$$\phi_e^a = -\sigma\epsilon^{ij}\partial_i\omega_j^a + \epsilon^{ij}\partial_i h_j^a + \frac{1}{2}\epsilon^{ij}\epsilon_{bc}^a\omega_i^b h_j^c - \frac{\sigma}{2}\epsilon^{ij}\epsilon_{bc}^a\omega_i^b\omega_j^c,$$

$$\phi_\omega^a = -\sigma\epsilon^{ij}\partial_i e_j^a + \frac{1}{\mu}\epsilon^{ij}\partial_i\omega_j^a - \frac{\sigma}{2}\epsilon^{ij}\epsilon_{bc}^a e_i^b\omega_j^c + \frac{1}{2}\epsilon^{ij}\epsilon_{bc}^a e_i^b h_j^c$$
$$+ \frac{1}{2\mu}\epsilon^{ij}\epsilon_{bc}^a\omega_i^b\omega_j^c,$$

$$\phi_h^a = \epsilon^{ij}\partial_i e_j^a + \frac{1}{2}\epsilon^{ij}\epsilon_{bc}^a e_i^b\omega_j^c. \quad (52)$$

In this way among 27 variables $e_\mu^a$, $\omega_\mu^a$ and $h_\mu^a$, the variables $e_0^a$, $\omega_0^a$ and $h_0^a$ are Lagrange multipliers and the remaining 18 variables $e_i^a$, $\omega_i^a$ and $h_i^a$ are dynamical. Using Eq. (41) for Poisson brackets and Eq. (49) for coefficients of the model, we find the following (nonvanishing) brackets

$$\{e_i^a(\mathbf{x}), h_j^b(\mathbf{x'})\} = \epsilon_{ij}\eta^{ab}\delta^2(\mathbf{x} - \mathbf{x'}),$$
$$\{\omega_i^a(\mathbf{x}), \omega_j^b(\mathbf{x'})\} = \mu\epsilon_{ij}\eta^{ab}\delta^2(\mathbf{x} - \mathbf{x'}),$$
$$\{\omega_i^a(\mathbf{x}), h_j^b(\mathbf{x'})\} = \sigma\mu\epsilon_{ij}\eta^{ab}\delta^2(\mathbf{x} - \mathbf{x'}),$$
$$\{h_i^a(\mathbf{x}), h_j^b(\mathbf{x'})\} = \sigma^2\mu\epsilon_{ij}\eta^{ab}\delta^2(\mathbf{x} - \mathbf{x'}). \quad (53)$$

Since the canonical Hamiltonian is zero for CSL theories, the total Hamiltonian density in Eq. (42) reads

$$\mathcal{H}_T = -e_0.\phi_e - \omega_0.\phi_\omega - h_0.\phi_h. \quad (54)$$

The matrix $M$ of Eq. (44) for the case of TMG contain nine $3 \times 3$ sub-matrices of the form $M_{e,e} = \{\phi_e, \phi_e\}$, $M_{e,\omega} = \{\phi_e, \phi_\omega\}$, etc. These are given in Appendix A. It comes out that the constraints $\phi_\omega^a$ have weakly vanishing Poisson brackets with each other and with the constraints $\phi_e^a$ and $\phi_h^a$. So the constraints $\phi_\omega^a$ are first class and the constraints $\phi_e^a$ and $\phi_h^a$ are second class. It makes the rank of $M$ equal to 6 (i.e., $m_0 = 6$), corresponding to 6 second class constraints. In the framework of our notations of Sec. II the constraints $\phi_\omega^a$ have the role of the constraints $\phi_A$ which generate the unlimited gauge transformations. The algebra of these first class constraints as given in Eqs. (A10) exhibits the algebra of the generators of the Lorentz group in $(2+1)$ dimensions. Explicit variations of the canonical variables under Poisson bracket with the generating functional constructed from $\phi_\omega$ (given in Appendix B), also agrees that the three first class constraints $\phi_\omega$ are in fact generators of the local Lorentz transformations (LLT). This symmetry is apparent from the tensor structure of the Lagrangian (48) with respect to the flat space indices, where all of the $a, b, \ldots$ indices are summed up.

To show these results explicitly, we give the final form of Eq. (43) for consistency of the constraints as follows:

$$-\{\phi_\omega^a, H_T\} = \omega_0^b \epsilon_{bc}^a \phi_\omega^c(x_\mu) + e_0^b \epsilon_{bc}^a \phi_e^c(x_\mu) + h_0^b \epsilon_{bc}^a \phi_h^c(x_\mu)$$
$$\approx 0, \quad (55)$$

$$-\{\phi_e^a, H_T\} = \omega_0^b \epsilon_{bc}^a \phi_e^c(x_\mu) - e_0^b \frac{\mu}{2} \epsilon^{ij} h_i^a h_j^b - h_0^b \frac{\mu}{2} \epsilon^{ij} e_i^a h_j^b$$
$$\approx 0, \quad (56)$$

$$-\{\phi_h^a, H_T\} = \omega_0^b \epsilon_{bc}^a \phi_h^c - e_0^b (\epsilon_{bc}^a m (\phi_\omega^c - \phi_e^c) - m\epsilon^{ij} e_i^a h_j^b)$$
$$+ h_0^b \left(\frac{\mu}{2} \epsilon^{ij} e_i^a e_j^b\right) \approx 0. \quad (57)$$

On the constraint surface all terms of Eq. (55) as well as the first terms of Eqs. (56) and (57) vanish; verifying that the constraints $\phi_\omega^a$ are first class. Hence, the Lagrange multipliers $\omega_0^a$ remain undetermined as arbitrary gauge parameters (LLT parameters, in fact). Let us then rewrite the consistency equations for the constraints $\phi_e^a$ and $\phi_h^a$ [i.e., Eqs. (56) and (57)], on the constraint surface, as follows

$$\begin{pmatrix} \frac{\mu}{2}\epsilon^{ij}h_i^a h_j^b & \frac{\mu}{2}\epsilon^{ij}e_i.h_j\delta^{ab} + \frac{\mu}{2}\epsilon^{ij}e_i^a h_j^b \\ -\frac{\mu}{2}\epsilon^{ij}e_i.h_j\delta^{ab} - \frac{\mu}{2}\epsilon^{ij}e_i^a h_j^b & -\frac{\mu}{2}\epsilon^{ij}e_i^a e_j^b \end{pmatrix}$$
$$\times \begin{pmatrix} e_0^b \\ h_0^b \end{pmatrix} \approx 0, \quad (58)$$

where the matrix elements of the $6 \times 6$ matrix on the left-hand side of Eq. (58) can be read directly from the expressions on the right-hand side of Eqs. (56) and (57). This matrix is the matrix $M^1$ in the terminology of Sec. II. Since $\phi_e^a$ and $\phi_h^a$ are second class, $M^1$ is full rank and, without imposing any further constraint, Eq. (58) implies that $e_0^a = 0$ and $h_0^a = 0$. Using Eq. (31) this gives $g_{00} = g_{0i} = 0$ which is just a spacial, though dynamical, metric. For a gravity model this result is not acceptable since it leads to a zero determinant metric. This seems as an unwanted dynamical branch of TMG, within the whole phase space of the variables. For this branch Eq. (21) gives the number of degrees of freedom as

$$D = 18 - 2 \times 3 - 6 = 6. \quad (59)$$

It is noticeable that if we omit the term $h.T$ (which guarantees the torsion-free condition) from the Lagrangian (48) of TMG, we would obtain six first class constraints $\phi_\omega^a$ and $\phi_e^a$ which generate both LLT and diffeomorphism. However, such a theory with 12 canonical variables ($e_i^a$ and $\omega_i^a$) would have no dynamical degree of freedom according to 6 first class constraints. This theory is, in fact, equivalent to original Hilbert-Einstein theory in three dimensions.

As mentioned generally in Sec. II, a dynamical system may have different behaviors in different subregions of the phase space where the algebra of the constraints may differ from the bulk. More precisely, Eq. (58), as the inevitable consequence of the dynamics of the action (48), can be satisfied in another way; i.e., instead of $e_0^a = 0$ and $h_0^a = 0$, we can assume the determinant of $M^1$ may vanish in some special region of the phase space, such that the matrix $M^1$ has some definite non trivial null vectors with nonvanishing components $e_0^a$. In other words, the rank of the matrix of Poisson brackets is not necessarily constant throughout the whole phase space.

This possibility, which is well known in the literature of constrained systems [26], should not be ignored in treating the consistency of constraints. However, this point has not been recognized as the origin of introducing new constraints, yet. In our current case, it is easy to see that the rank of matrix $M^1$ reduces, for instance, for the flat metric solution given by Eqs. (71) below. In fact, this is also the case for small perturbations around the flat solution, as we will explicitly show at the end of this section.

To find the new constraints, which describe the physical sector of the phase space, we should focus on the consistency Equations (58). These are six equations for six unknown $e_0^a$ and $h_0^a$. Following the general instructions of Sec. II, we should find the vanishable factors of $\det M^1$. Since the elements of $M^1$ are quadratic function of $e_i^a$ and $h_i^a$, $\det M^1$ is a polynomial of order 12 with respect to these variables. It is very difficult to find different factors of $\det M^1$ directly. However, we may use a trick. Let us rewrite the Eq. (58) in the following form

$$\begin{pmatrix} H & T+X \\ -T-X & -E \end{pmatrix} \begin{pmatrix} e_0^b \\ h_0^b \end{pmatrix} \approx 0, \quad (60)$$

where

$$H \equiv \frac{\mu}{2}\epsilon^{ij}h_i^a h_j^b, \quad E \equiv \frac{\mu}{2}\epsilon^{ij}e_i^a e_j^b, \quad T \equiv \frac{\mu}{2}\epsilon^{ij}e_i^a h_j^b,$$
$$X \equiv \frac{\mu}{2}\epsilon^{ij}e_i.h_j\delta^{ab}, \quad (61)$$

are invertible $3 \times 3$ matrices. We can multiply both sides of Eq. (60) by the arbitrary matrix

$$\begin{pmatrix} A & B \\ C & D \end{pmatrix}. \quad (62)$$

The choice

$$A = X(H - (X+T)E^{-1}(X+T))^{-1},$$
$$B = X(H - (X+T)E^{-1}(X+T))^{-1}(X+T)E^{-1},$$
$$C = 1, \quad D = 1, \quad (63)$$

converts Eq. (60) to the following form

$$\begin{pmatrix} X & 0 \\ H-X-T & X+T-E \end{pmatrix} \begin{pmatrix} e_0^b \\ h_0^b \end{pmatrix} \approx 0. \qquad (64)$$

Considering the first row of the Eq. (64) gives

$$X e_0^a = \frac{\mu}{2} \epsilon^{ij} e_i . h_j \delta^{ab} e_0^b = \frac{\mu}{2} \epsilon^{ij} e_i . h_j e_0^a \approx 0. \qquad (65)$$

which can be written as $(\Gamma \delta_a^b) e_0^a = 0$ where

$$\Gamma \equiv \epsilon^{ij} e_i . h_j = \epsilon^{ij} \eta_{ab} e_i^a h_j^b \qquad (66)$$

Note also that the Eq. (65) can be deduced from Eq. (46) by suitable choice of indices. The only possibility for satisfying Eq. (65) for nonvanishing $e_0^a$ is assuming $\Gamma$ as a new constraint. We emphasis that $\Gamma \approx 0$ is not a necessary consequence of the original dynamics of the theory. The equations of motion, in any way written, do not give more than Eq. (65). Therefore, one is not able to consider Eq. (66) as a secondary constraint.

On the other hand, if we denote the matrix on the left-hand side of Eq. (64) as $\tilde{M}^1$ then it is apparent that $\det \tilde{M}^1$ is proportional to $\det M^1$. Fortunately $\tilde{M}^1$ has a vanishing upright block and we can write

$$\det M^1 \propto \det \tilde{M}^1 = \det X \det(X+T-E). \qquad (67)$$

In other words, it is apparent that $\det \tilde{M}^1$ should at least have a factor of $\Gamma^3$. However, we know from Sec. II that every vanishable factor in $\det M^1$ should be of even order. Hence $\det M^1$ should contain a factor of $\Gamma^4$. In other words, the rank of $M^1$ reduce suddenly from 6 to 2 on the surface $\Gamma = 0$ i.e., $\tilde{m} = 2$. Hence, using the general result (45) for CLS theories we find $D = 2$ for the physical sector of TMG, which is equivalent to one degree of freedom in configuration space.

The main point in the reduction the number of degrees of freedom from 6 in Eq. (59) to 2, is converting 3 second class constraints to first class, at the expense of introducing one more second class constraint (i.e., $\Gamma$) to the system. This procedure decrease $3+1$ degrees of freedom.

Emerging 3 first class constraints on the surface $\Gamma = 0$, is a good news for the symmetries of the system. In fact, we expected in advanced three first class constraints as the generators of diffeomorphism in the framework of vielbein formalism. Fortunately in Ref. [25] it is explicitly shown that for infinitesimal diffeomorphism $x^\mu \to x^\mu - \xi^\mu$ the gauge generator $G = \xi^\mu \psi_\mu$ gives the variation of every function $f$ as $\delta f = \{f, G\}$ where

$$\psi_\mu = e_\mu . \phi_e + h_\mu . \phi_h + \omega_\mu . \phi_\omega. \qquad (68)$$

The important point is that the generators $\psi_\mu$ are not originally first class constraints, unless we land on the surface $\Gamma = 0$. However, before doing that, putting the first class constraints $\phi_\omega$ away, we can change the set of second class constraints from 6 constraints $\phi_e$ and $\phi_h$ to 6 (already second class) constraints $\psi$ and (say) $\phi_h$. Using Eq. (A10) of Appendix A, the Poisson brackets among $\psi_\mu$ and $\phi_h^a$ are

$$\{\phi_h^a, \phi_h^b\} = \epsilon^{ij} e_i^b e_j^a,$$
$$\{\psi_\mu, \psi_\nu\} = (e.h)(e_\mu . h_\nu - e_\nu . h_\mu),$$
$$\{\psi_\mu, \phi_h^b\} = -(e.h) e_\mu^b. \qquad (69)$$

It is obvious that in the subregion specified by $e.h = 0$, the constraints $\psi_\mu$ convert to first class. Let denote the matrix of Poisson brackets of the constraints $\psi_\mu$ and $\phi_h$ as $M'^1$. One can show that under redefinition of the constraints the matrix of Poisson brackets would be proportional to the previous one. From Eqs. (69) it is easy to see that $\det M'^1$ is proportional to $\Gamma^4$. This is, in fact, a further proof of our assertion above that $\det M^1 \propto \Gamma^4$. As mentioned in Sec. II it is also necessary to investigate consistency of the new constraint $\Gamma$ under the evolution of the system in time. In other words, we need to assure about validity of the equation $\{\Gamma, H_T\} \approx 0$, where $H_T$ is given in Eq. (54).

Direct calculation shows that $\Gamma$ commutes with the first class constraints $\phi_w$, as expected from our claim in Sec. II. This is a good news, since we expect that the already established unlimited gauge transformations (i.e., LLT) are not distorted on the physical sector of the theory. On the other hand, $\Gamma$ does not commute with the constraints $\psi_\mu$ as well as $\phi_h$. However, this dose not mean that the constraint $\Gamma$ makes the whole system of constraints $\psi_\mu$ and $\phi_h$ second class. It is well known [14] that one individual second class constraint can be conjugated with only one constraint. On the other hand, remembering that the rank of $M'^1$ is two, means that two existing constraints, say two of the constraints $\phi_h$ is already second class. The third one can be considered as the conjugate constraint to $\Gamma$. Hence, we can consider three constraints $\phi_h$ and the constraint $\Gamma$ as a system of 4 second class constraints. Assuming the extended Hamiltonian for the whole set of constraints as

$$\mathcal{H}_E = \xi^\mu \psi_\mu + \omega_0 . \phi_\omega + h_0 . \phi_h + \lambda \Gamma, \qquad (70)$$

it is easy to check that the consistency of the constraints can fix 4 Lagrange multipliers $h_0^a$ and $\lambda$, while six Lagrange multipliers $\omega_0^a$ and $\xi^\mu$ remain undetermined as the free parameters of LLT and diffeomorphism, respectively.

### A. Linearized TMG

Similar to so many models presented as extension of GR, the dynamical content of TMG, has been considered mostly in the linearized version of the model. The well-known result of this procedure is that the TMG in the linearized version contain only one degree of freedom in the metric formalism, which is a massive scaler with mass $\mu$ [3]. In this subsection we show explicitly how the dynamical structure of the theory behaves after linearization.

In order to linearize the model we need a background solution where the dynamical variables will be small perturbation around that background. For the case of TMG with zero cosmological constant the flat space-time can survive as a background solution. One may easily check that the solution

$$e_\mu^a = \delta_\mu^a, \qquad \omega_\mu^a = 0, \qquad h_\mu^a = 0, \qquad (71)$$

which leads to the flat metric, satisfies the equations of motion (50). Let us denote fluctuations of one-forms around the background (71) as $\zeta_\mu^a$, $\omega_\mu^a$ and $h_\mu^a$ respectively, where

$$e_\mu^a = \delta_\mu^a + \zeta_\mu^a. \qquad (72)$$

Keeping up to quadratic terms with respect to the excitations, the linearized Lagrangian density reads

$$\mathcal{L}_{\text{lin}} = \frac{\sigma}{2}\varepsilon^{ij}\zeta_i^a\dot{\omega}_j^a + \frac{\sigma}{2}\varepsilon^{ij}\omega_i^a\dot{\zeta}_j^a - \frac{1}{2\mu}\varepsilon^{ij}\omega_i^a\dot{\omega}_j^a - \frac{1}{2}\varepsilon^{ij}h_i.\dot{\zeta}_j$$
$$- \frac{1}{2}\varepsilon^{ij}\zeta_i.\dot{h}_j + \zeta_0^a\psi_\zeta^a + \omega_0^a\psi_\omega^a + h_0^a\psi_h^a - \frac{\sigma}{2}\varepsilon^{ij}\varepsilon_{bc}\omega_i^b\omega_j^c$$
$$+ \frac{1}{2}\varepsilon^{ij}\varepsilon_{bc}h_i^b\omega_j^c, \qquad (73)$$

where $\varepsilon_{bc} \equiv \varepsilon_{0bc}$ and $\psi$'s are

$$\psi_\zeta^a = \varepsilon^{ij}(-\sigma\partial_i\omega_j^a + \partial_i h_j^a),$$
$$\psi_\omega^a = \varepsilon^{ij}\left(-\sigma\partial_i\zeta_j^a + \frac{1}{\mu}\partial_i\omega_j^a - \frac{\sigma}{2}\varepsilon_{bc}^a\delta_i^b\omega_j^c + \frac{1}{2}\varepsilon_{bc}^a\delta_i^b h_j^c\right),$$
$$\psi_h^a = \varepsilon^{ij}\left(\partial_i\zeta_j^a + \frac{1}{2}\varepsilon_{bc}^a\delta_i^b\omega_j^c\right). \qquad (74)$$

As is seen, the linearized model is almost a different theory whose solutions coincide with a limited class of the solutions of the original theory, i.e., small perturbation around the flat background. In contrast to the original model, in the linearized version of TMG the canonical Hamiltonian no longer vanishes; instead we have

$$\mathcal{H}_C = -\frac{\sigma}{2}\varepsilon^{ij}\varepsilon_{bc}\omega_i^b\omega_j^c + \frac{1}{2}\varepsilon^{ij}\varepsilon_{bc}h_i^b\omega_j^c. \qquad (75)$$

As before, the variables $\zeta_0^a$, $\omega_0^a$, and $h_0^a$ behave as Lagrange multipliers, while the new functions $\psi_\zeta^a$, $\psi_\omega^a$, and $\psi_h^a$ behave as the constraints of the linearized model.

Fortunately the kinetic terms of the original Lagrangian is linear in advanced; so the symplectic matrix, as well as the fundamental Poisson brackets do not change at all, i.e., the relations (53) is valid with $e_\mu^a$ replaced by $\zeta_\mu^a$. The total Hamiltonian is $H_T = \int d^2\mathbf{x}\mathcal{H}_T$, where

$$\mathcal{H}_T = -\zeta_0.\psi_\zeta - \omega_0.\psi_\omega - h_0.\psi_h + \mathcal{H}_C. \qquad (76)$$

Consistency of the constraints $\psi_\zeta^a$ and $\psi_\omega^a$ is satisfied identically, i.e.,

$$\{\psi_\zeta^a, H_T\} \approx 0,$$
$$\{\psi_\omega^a, H_T\} \approx 0, \qquad (77)$$

which shows that the 6 constraint $\psi_\zeta^a$ and $\psi_\omega^a$ are first class. Consistency of the constraints $\psi_h^a$ has different results for $a = 0, 1, 2$, as follows

$$\{\psi_h^0, H_T\} = h_1^2 - h_2^1, \qquad (78)$$

$$\{\phi_h^1, H_T\} = \partial_1\omega_2^2 - \partial_2\omega_1^2 - \frac{\mu}{2}h_0^2, \qquad (79)$$

$$\{\psi_h^2, H_T\} = \partial_2\omega_1^1 - \partial_1\omega_2^1 - \frac{\mu}{2}h_0^1. \qquad (80)$$

Note that the right-hand side of Eq. (78), as well as the first two terms on the right-hand side of Eqs. (79) and (80), have emerged due to the existence of a canonical Hamiltonian for linearized model. Equations (79) and (80) can be used ro determine the Lagrange multipliers $h_0^1$ and $h_0^2$, while the Eq. (78) should be considered as a new constraint $\Omega = h_1^2 - h_2^1$. According to the Dirac prescription for the constrained systems, we should investigate consistency of the new constraint $\Omega$ as follows

$$\{\Omega, H_T\} = \frac{-\sigma m^2}{2}(h_1^1 + h_2^2) + \sigma m^2 h_0^0 + m\partial_2 h_0^1 - m\partial_1 h_0^2. \qquad (81)$$

Since the Lagrange multipliers $h_0^1$ and $h_0^2$ have been determined in the previous step, the remaining Lagrange multiplier $h_0^0$ is determined by the Eq. (81).

In this way the 6 Lagrange multipliers $\zeta_0^a$ and $\omega_0^a$ remain undetermined and three Lagrange multipliers $h_0^a$ are determined in the consistency process. Considering the constraint structure, we have 10 constraints for the linearized theory, i.e., one more constraint $\Omega$ has arisen due to different canonical structure of the linearized model. The first class constraints $\psi_\zeta^a$ and $\psi_\omega^a$ remain first class during the process of consistency. Hence, we have all together 6 first class and 4 second class constraints on 18 canonical variable. Using the basic formula (21) for the number of degree of freedom we have

$$18 - 6 \times 2 - 4 \times 1 = 2, \qquad (82)$$

which correspond to one degree of freedom in the configuration space. This is the desired result in agreement with what obtained in Ref. [3] in metric formalism.

It is interesting that primary constraints $\psi$ given in Eqs. (74) in linear analysis are the linearized version of the primary constraint $\phi$ in Eqs. (52) around flat metric. Moreover, it is noticeable that the secondary constraint $\Omega$ is the same new constraint $e.h$ linearized around flat metric.

## V. HAMILTONIAN STRUCTURE OF NMG

The Lagrangian of NMG in the framework of veilbein formalism is proposed as [17]

$$L = -\sigma e.R(\omega) + h.T(\omega) - \frac{1}{m^2} f.R(\omega) - \frac{1}{2m^2} f.e \times f. \tag{83}$$

The first term is equivalent to the scaler curvature of the Hilbert-Einstein gravity. The second term guarantees the torsion-free condition, and the new field $f_\mu^a$ acts as an auxiliary field. Using the equations of motion of the auxiliary field, one can write it in terms of the original fields $e_\mu^a$ and $\omega_\mu^a$. This can be shown to lead to the original form of the term $(R^{\mu\nu}R_{\mu\nu} - \frac{3}{8}R^2)$ of NMG in the metric formalism. Hence, in NMG we have altogether four flavors $e$, $\omega$, $h$, and $f$. Comparing to the general form of the Chern-simons like Lagrangian (35), the coefficients $g_{rs}$ and $f_{rst}$ are as follows

$$g_{e\omega} = -\sigma \quad g_{eh} = 1 \quad g_{f\omega} = -\frac{1}{m^2}$$

$$f_{e\omega\omega} = -\sigma \quad f_{eh\omega} = 1 \quad f_{f\omega\omega} = -\frac{1}{m^2}$$

$$f_{eff} = -\frac{1}{m^2}. \tag{84}$$

The equations of motion for the Lagrangian (83) are as follows

$$T(\omega) = 0,$$
$$R(\omega) + e \times f = 0,$$
$$df + \omega \times f - m^2(e \times h) = 0,$$
$$dh + \omega \times h - \sigma R(\omega) - \frac{1}{2m^2} f \times f = 0. \tag{85}$$

The Lagrangian density corresponding to the Lagrangian (83) can be written as

$$\mathcal{L} = \frac{\sigma}{2} \epsilon^{ij} e_i.\dot\omega_j + \frac{\sigma}{2} \epsilon^{ij}\omega_i.\dot e_j - \frac{1}{2}\epsilon^{ij} h_i.\dot e_j - \frac{1}{2}\epsilon^{ij} e_i.\dot h_j$$
$$- \frac{1}{m^2}\epsilon^{ij}\omega_i.\dot f_j - \frac{1}{m^2}\epsilon^{ij} f_i.\dot\omega_j + e_0.\phi_e + \omega_0.\phi_\omega$$
$$+ h_0.\phi_h + f_0.\phi_f, \tag{86}$$

where 12 primary constraints of the system read

$$\phi_h^a = \epsilon^{ij}\partial_i e_j^a + \frac{1}{2}\epsilon^{ij}\epsilon_{bc}^a e_i^b \omega_j^c,$$

$$\phi_f^a = -\frac{1}{m^2}\epsilon^{ij}\partial_i \omega_j^a - \frac{1}{2m^2}\epsilon^{ij}\epsilon_{bc}^a \omega_i^b \omega_j^c - \frac{1}{m^2}\epsilon^{ij}\epsilon_{bc}^a e_i^b f_j^c,$$

$$\phi_e^a = -\sigma\epsilon^{ij}\partial_i \omega_j^a + \epsilon^{ij}\partial_i h_j^a - \frac{\sigma}{2}\epsilon^{ij}\epsilon_{bc}^a \omega_i^b \omega_j^c + \frac{1}{2}\epsilon^{ij}\epsilon_{bc}^a h_i^b \omega_j^c$$
$$- \frac{1}{2m^2}\epsilon^{ij}\epsilon_{bc}^a f_i^b f_j^c,$$

$$\phi_\omega^a = -\sigma\epsilon^{ij}\partial_i e_j^a - \frac{1}{m^2}\epsilon^{ij}\partial_i f_j^a - \frac{\sigma}{2}\epsilon^{ij}\epsilon_{bc}^a e_i^b \omega_j^c + \frac{1}{2}\epsilon^{ij}\epsilon_{bc}^a e_i^b h_j^c$$
$$- \frac{1}{2m^2}\epsilon^{ij}\epsilon_{bc}^a \omega_i^b f_j^c. \tag{87}$$

The model possesses 12 Lagrange multipliers $e_0^a$, $\omega_0^a$, $h_0^a$, and $f_0^a$ and 24 canonical variables $e_i^a$, $\omega_i^a$, $h_i^a$ and $f_i^a$. According to the general form of the Poisson brackets given in Eq. (41) the kinetic term in the Lagrangian density (86) determines the nonvanishing fundamental Poisson brackets as follow

$$\{e_i^a(\mathbf{x}), h_j^b(\mathbf{x}')\} = \epsilon_{ij}\eta^{ab}\delta^2(\mathbf{x}-\mathbf{x}'),$$
$$\{\omega_i^a(\mathbf{x}), f_j^b(\mathbf{x}')\} = -m^2\epsilon_{ij}\eta^{ab}\delta^2(\mathbf{x}-\mathbf{x}'),$$
$$\{h_i^a(\mathbf{x}), f_j^b(\mathbf{x}')\} = -\sigma m^2 \epsilon_{ij}\eta^{ab}\delta^2(\mathbf{x}-\mathbf{x}'). \tag{88}$$

The total Hamiltonian can be read from Eq. (86) as

$$\mathcal{H}_T = -e_0.\phi_e - \omega_0.\phi_\omega - h_0.\phi_h - f_0.\phi_f. \tag{89}$$

Using the algebra of primary constraints derived in the Appendix A, consistency of the constraints (87) is performed as follows

$$-\{\phi_\omega^a, H_T\} = \omega_0^b \epsilon_{bc}^a \phi_\omega^c + e_0^b \epsilon_{bc}^a \phi_e^c + h_0^b \epsilon_{bc}^a \phi_h^c + f_0^b \epsilon_{bc}^a \phi_f^c \approx 0, \tag{90}$$

$$-\{\phi_e^a, H_T\}$$
$$= \omega_0^b \epsilon_{bc}^a \phi_e^c + e_0^b \left[\frac{\mu}{2}\epsilon^{ij} f_i^a h_j^b - \frac{\mu}{2}\epsilon^{ij} f_i^b h_j^a\right]$$
$$+ h_0^b \left[\epsilon_{bc}^a \epsilon^{ij}\partial_i \omega_j^c - \frac{\epsilon^{ij}}{2} f_i^c e_j^c \delta^{ab} - \frac{\epsilon^{ij}}{2}\omega_i^a\omega_j^b - \frac{\epsilon^{ij}}{2} e_i^a f_j^b\right]$$
$$+ f_0^b \left[\epsilon_{bc}^a \epsilon^{ij}\partial_i f_j^c + \frac{\epsilon^{ij}}{2} e_i^c h_j^c \delta^{ab} + \frac{\epsilon^{ij}}{2m^2} f_i^a \omega_j^b - \frac{\epsilon^{ij}}{2m^2} f_i^b \omega_j^a\right]$$
$$- \frac{1}{2}\epsilon^{ij} e_i^b h_j^a\right] \approx 0, \tag{91}$$

$$-\{\phi_h^a, H_T\} = \omega_0^b \epsilon_{bc}^a \phi_h^c + e_0^b \left[\epsilon_{bc}^a \epsilon^{ij}\partial_i \omega_j^c - \frac{\epsilon^{ij}}{2} f_i^c e_j^c \delta^{ab} - \frac{\epsilon^{ij}}{2}\omega_i^a\omega_j^b - \frac{\epsilon^{ij}}{2} e_i^a f_j^b\right] + f_0^b \frac{\mu}{2}\epsilon^{ij} e_i^a e_j^b \approx 0, \tag{92}$$

$$-\{\phi_f^a, H_T\} = \omega_0^b \epsilon_{bc}^a \phi_f^c + e_0^b \left[ \epsilon_{bc}^a \epsilon^{ij} \partial_i f_j^c + \frac{\epsilon^{ij}}{2} e_i^c h_j^c \delta^{ab} \right.$$
$$\left. + \frac{\epsilon^{ij}}{2m^2} f_i^a \omega_j^b - \frac{\epsilon^{ij}}{2m^2} f_i^b \omega_j^a - \frac{1}{2} \epsilon^{ij} e_i^b h_j^a \right]$$
$$+ f_0^b [\epsilon_{bc}^a \psi_h^c] + h_0^b \left[ \frac{\mu}{2} \epsilon^{ij} e_i^a e_j^b \right] \approx 0. \quad (93)$$

Equation (90) requires that three constraints $\phi_\omega^a$ are first class. As in the case of TMG, it can be shown directly that these three constraints satisfy the algebra of Lorentz group in three dimensions. They also generate, under taking the Poisson brackets, the LLT on the variables of the theory (see Appendix B). As before, the LLT are the unlimited gauge transformations of the theory.

Since coefficients of the Lagrange multipliers $\omega_0^a$ in Eqs. (91), (92), and (93) vanish weakly, these equations can be viewed as 9 equations to determine 9 Lagrange multipliers $e_0^b$, $h_0^b$ and $f_0^b$ as follows

$$\begin{pmatrix} \{\phi_e, \phi_e\} & \{\phi_e, \phi_h\} & \{\phi_e, \phi_f\} \\ -\{\phi_e, \phi_h\} & \{\phi_h, \phi_h\} & \{\phi_h, \phi_f\} \\ -\{\phi_e, \phi_f\} & -\{\phi_h, \phi_f\} & \{\phi_f, \phi_f\} \end{pmatrix} \begin{pmatrix} e_0 \\ h_0 \\ f_0 \end{pmatrix} \approx 0. \quad (94)$$

The matrix on the right-hand side of Eq. (94) is the matrix $M^1$ discussed in Sec. II where its elements can be read from Eqs. (91), (92), and (93). The main problem is to determine the rank of this matrix on the physical sector of the theory where the veilbein variables are invertible and supports the existence of appropriate first class constraints as the generators of diffeomorphism. Our experience in TMG shows that it is very difficult to calculate directly the determinant of the matrix $M^1$ and find its vanishable factors. However, we can rely on our physical expectation that on the physical sector we should have first class constraints which generate diffeomorphism. The generators of diffeomorphism can be suggested as the generalization of the Carlip formula [25], for TMG given in Eq. (68) as

$$\psi_\mu = e_\mu . \phi_e + h_\mu . \phi_h + \omega_\mu . \phi_\omega + f_\mu . \phi_f. \quad (95)$$

We can replace three constraints $\phi_e$ with three constraints $\psi_\mu$ and consider the set of 9 constraints $\phi_h^a$, $\phi_f^a$ and $\psi_\mu$ with the following Poisson brackets (derived from Appendix A)

$$\{\phi_h^a, \phi_h^b\} = 0$$
$$\{\phi_f^a, \phi_f^b\} \approx \epsilon_c^{ab} \phi_h^c,$$
$$\{\phi_h^a, \phi_f^b\} \approx \epsilon^{ij} e_i^b e_j^a,$$
$$\{\psi_\mu, \psi_\nu\} \approx (\epsilon^{ij} e_i . h_j)(e_\mu . f_\nu - e_\nu . f_\mu)$$
$$\qquad + (\epsilon^{ij} e_i . f_j)(e_\mu . h_\nu - e_\nu . h_\mu),$$
$$\{\psi_\mu, \phi_h^b\} \approx (\epsilon^{ij} e_i . f_j) e_\mu^b,$$
$$\{\psi_\mu, \phi_f^b\} \approx (\epsilon^{ij} e_i . h_j) e_\mu^b. \quad (96)$$

They form the matrix $M'^1$ instead of the matrix $M^1$ of Eq. (94) as follows

$$\begin{pmatrix} \{\psi, \psi\} & \{\psi, \phi_h\} & \{\psi, \phi_f\} \\ -\{\psi, \phi_h\} & \{\phi_h, \phi_h\} & \{\phi_h, \phi_f\} \\ -\{\psi, \phi_f\} & -\{\phi_h, \phi_f\} & \{\phi_f, \phi_f\} \end{pmatrix} \begin{pmatrix} \xi \\ h_0 \\ f_0 \end{pmatrix} \approx 0. \quad (97)$$

where $\xi_\mu$ are the parameters of the infinitesimal diffeomorphism $x^\mu \to x^\mu - \xi^\mu$. From Eqs. (96) it is easy to see that the constraints $\psi_\mu$ are first class on the surface of the new constraints $\Gamma$ and $\Omega$, where

$$\Gamma \equiv \epsilon^{ij}(e_i . h_j) \approx 0, \qquad \Omega \equiv \epsilon^{ij}(e_i . f_j) \approx 0. \quad (98)$$

It can be seen directly from Eqs. (96) that under imposing the constraints (98) on the system rank of matrix $\tilde{M} \equiv M|_{e.h=0, e.f=0}$ would be equal to four, i.e., $\tilde{m} = 4$. Using the general result (45), this means that in the physical sector of the theory, NMG has four phase space degrees of freedom which corresponds to two degrees of freedom in configuration space. This result is consistent with the known results of the literature.

For the sake of completeness, let us consider the problem of consistency of the new constraints. Similar to TMG, we can add the new constraints to the total Hamiltonian to find the extended Hamiltonian as

$$\mathcal{H}_E = \xi^\mu \psi_\mu + \omega_0 . \phi_\omega + h_0 . \phi_h + f_0 . \phi_f + \lambda \Gamma + \eta \Omega. \quad (99)$$

The constraints $\Gamma$ and $\Omega$ do not commute with all of the constraints. Remember that, before considering the consistency of the new constraints, four constraints among $\phi_f$ and $\phi_h$ are second class and two are first class (on the surface of the constraints). Consistency of the new constraints makes these two constraints again second class. Hence, the set of 8 constraints $\Gamma$, $\Omega$, $\phi_f^a$, and $\phi_h^a$ constitute a second class system of constraints. In this way, using Eq. (21) the number of dynamical degrees of freedom reads

$$24 - 6 \times 2 - 8 \times 1 = 4, \quad (100)$$

as expected.

The new constraints (98) may be derived alternatively using the general formula (46) by inserting the structure coefficients $f_{\text{rst}}$ as follows

$$e^a e . h = 0, \qquad e^a e . f = 0. \quad (101)$$

By suitable choice of curved space indices and by assuming invertibility of $e_\mu^a$ Eqs. (101) leads to the constraints (98).

### A. Linearized model

Similar to the previous case, the flat metric solution given by

$$e_\mu^a = \delta_\mu^a, \quad \omega_\mu^a = 0, \quad h_\mu^a = 0, \quad f_\mu^a = 0, \quad (102)$$

can be considered as a background solution for the equations of motion (85). Assuming $e_\mu^a = \delta_\mu^a + \zeta_\mu^a$ and considering $\zeta_\mu^a$, $\omega_\mu^a$, $h_\mu^a$ and $f_\mu^a$ as small perturbations, the linearized Lagrangian density reads

$$\mathcal{L}_{\text{lin}} = \frac{\sigma}{2}\varepsilon^{ij}\zeta_i.\dot\omega_j + \frac{\sigma}{2}\varepsilon^{ij}\omega_i.\dot\zeta_j - \frac{1}{2}\varepsilon^{ij}h_i.\dot\zeta_j - \frac{1}{2}\varepsilon^{ij}\zeta_i.\dot h_j$$
$$- \frac{1}{2}\varepsilon^{ij}f_i.\dot\zeta_j - \frac{1}{2}\varepsilon^{ij}\zeta_i.\dot f_j + f_0.\psi_f + \zeta_0.\psi_\zeta$$
$$+ \omega_0.\psi_{.\omega} + h_0.\psi_h - \frac{\sigma}{2}\varepsilon^{ij}\varepsilon_{bc}\omega_i^b\omega_j^c + \frac{1}{2}\varepsilon^{ij}\varepsilon_{bc}h_i^b\omega_j^c$$
$$+ \frac{1}{2}\varepsilon^{ij}\varepsilon_{bc}f_i^b f_j^c, \quad (103)$$

where the primary constraints are

$$\psi_\zeta^a = \varepsilon^{ij}(-\sigma\partial_i\omega_j^a + \partial_i h_j^a),$$
$$\psi_\omega^a = \varepsilon^{ij}\left(-\sigma\partial_i\zeta_j^a + \frac{1}{m^2}\partial_i f_j^a - \frac{\sigma}{2}\epsilon_{bc}^a\delta_i^b\omega_j^c + \frac{1}{2}\epsilon_{bc}^a\delta_i^b h_j^c\right),$$
$$\psi_h^a = \varepsilon^{ij}\left(\partial_i\zeta_j^a + \frac{1}{2}\epsilon_{bc}^a\delta_i^b\omega_j^c\right),$$
$$\psi_f^a = \varepsilon^{ij}\left(-\partial_i\omega_j^a - \frac{1}{2}\epsilon_{bc}^a\delta_i^b f_j^c\right), \quad (104)$$

and the density of total Hamiltonian reads

$$\mathcal{H}_T = -f_0.\psi_f - \zeta_0.\psi_\zeta - \omega_0.\psi_{.\omega} - h_0.\psi_h + \frac{\sigma}{2}\varepsilon^{ij}\varepsilon_{bc}\omega_i^b\omega_j^c$$
$$- \frac{1}{2}\varepsilon^{ij}\varepsilon_{bc}h_i^b\omega_j^c + \frac{1}{2m^2}\varepsilon^{ij}\varepsilon_{bc}f_i^b f_j^c. \quad (105)$$

Using the fundamental Poisson bracket (88), with $e_\mu^a$ replaced by $\zeta_\mu^a$, we should evaluate consistency of the primary constraints. The constraints $\psi_\zeta^a$ and $\psi_\omega^a$ have vanishing Poisson brackets with the Hamiltonian, i.e.,

$$\{\psi_\zeta^a, H_T\} \approx 0,$$
$$\{\psi_\omega^a, H_T\} \approx 0. \quad (106)$$

This implicates that $\psi_\zeta^a$ and $\psi_\omega^a$ are first class constraints. The consistency of $\psi_h^2$, $\psi_h^3$, $\psi_f^2$, and $\psi_f^3$ eventuate to

$$\{\psi_h^1, H_T\} = \partial_1\omega_2^2 - \partial_2\omega_1^2 + \frac{1}{2}mf_0^2 \approx 0,$$
$$\{\psi_h^2, H_T\} = \partial_2\omega_1^1 - \partial_1\omega_2^1 + \frac{1}{2}mf_0^1 \approx 0,$$
$$\{\psi_f^1, H_T\} = \partial_1 f_2^2 - \partial_2 f_1^2 + \frac{1}{2}mh_0^2 \approx 0,$$
$$\{\psi_f^2, H_T\} = \partial_2 f_1^1 - \partial_1 f_2^1 + \frac{1}{2}mh_0^1 \approx 0, \quad (107)$$

which determine the Lagrange multipliers $h_0^1$, $h_0^2$, $f_0^1$, $f_0^2$, respectively. This indicates that the constraints $\psi_h^1$, $\psi_h^2$, $\psi_f^1$, and $\psi_f^2$ are second class. Consistency of the constraints $\psi_f^0$ and $\psi_h^0$ produce two new constraints as follows

$$\Sigma_1 \equiv \{\psi_h^0, H_T\} = h_1^2 - h_2^1,$$
$$\Sigma_2 \equiv \{\psi_f^0, H_T\} = f_1^2 - f_2^1. \quad (108)$$

For consistency of these new constraints we have

$$\{\Sigma_1, H_T\} = -\sigma(f_1^1 + f_2^2) + 2\sigma f_0^0 + 2\partial_2 h_0^1 - m\partial_1 h_0^2,$$
$$\{\Sigma_2, H_T\} = m^2(h_1^1 + h_2^2) + 2\sigma h_0^0 + 2\partial_2 f_0^1 - m\partial_1 f_0^2. \quad (109)$$

Equations (109) determine $h_0^0$ and $f_0^0$, which means that the constraints $\Sigma_1$ and $\Sigma_2$ as well as their parents $\psi_f^0$ and $\psi_h^0$ are second class. Therefore, the theory of NMG in the linear limit has 6 first class and 8 second class constraints. Using the formula (21), we can count the number of degrees of freedom in phase space as

$$24 - 6 \times 2 - 8 \times 1 = 4, \quad (110)$$

which corresponds to 2 degrees of freedom in the configuration space.

Again one can verify that primary constraints in the linear analysis are the linearized versions of the original primary constraints. The secondary constraints $\Sigma_1$ and $\Sigma_2$ are also linearized version of the new constraints $e.h$ and $e.f$. In the linearized model we see very clearly that the 4 constraints $\psi_h^2$, $\psi_h^3$, $\psi_f^2$, and $\psi_f^3$ are originally second class [see Eq. (107)], while the remaining two constraints $\psi_f^0$ and $\psi_h^0$ make a second class system with the secondary constraints $\Sigma_1$ and $\Sigma_2$.

## VI. HAMILTONIAN STRUCTURE OF ZDG

Multimetric gravity theories are generalizations of massive gravity which have gained considerable attention recently [27–29]. In three dimensions multimetric theories in the framework of veilbein formalism can be considered as CSL theories. Here we concentrate on Zwei-Dreibein gravity (ZDG) proposed in [10] which deals about two interacting vielbeins $e_\mu^a$ and $é_\mu^a$ and theirs associated spin-connections $\omega_\mu^a$ and $\omega'^a_\mu$, respectively. Hence, in ZDG we have four flavors.

Absence of the Boulware-deser (BD) ghost [30], like every other gravitational theory, is an essential point in multimetric gravity and specially in ZDG. Under linearization of the theory around the AdS solution [10], the Lagrangian density divides into a massless and a massive Pauli-Fierz divisions. It is well known that ordinary massless gravity has no dynamical degree of freedom in three dimensions, while the massive gravity has two degrees of freedom. Hence, the linearized ZDG possesses two degrees of freedom. Our experience about TMG and NMG showed that the number of degrees of freedom in the physical sector of the non linearized theory is the same as what found in the linearized version of the theory. Hence, the Boulware-Deser ghost would be absent if the Hamiltonian analysis of the theory supports this result, i.e., two dynamical degrees of freedom in the physical sector of the theory.

As stated in Sec. III for CSL theories, for four flavors of ZDG we have 24 canonical variables and 12 primary constraints. Based on this fact, the authors of Ref. [11] argued that in order to generate LLT and diffeomorphism one needs to have 6 first class constraints. According to this analysis there remain 6 second class constraints and the number of degrees of freedom using the formula (21) reads

$$24 - 6 \times 2 - 6 \times 1 = 6, \quad (111)$$

which is equivalent to 3 degrees of freedom in configuration space. Hence, Ref. [11] deduced that the ZDG theory has goast. However, they do not specify which constraints are generators of LLT and which are generators of diffeomorphism.

In response to this opposition, authors of Ref. [10] pointed to two more constraints which emerge by manipulating the equations of motion similar to what stated generally in Eqs. (46) and (47). These two second class constraints solve the problem of one more degrees of freedom. However, the exact algebra of the constraints and the generators of symmetries are not distinguished in the existing literature. Moreover, the exact way of emerging the new constraints in the canonical treatment of the theory as indicating the physical subregion of the phase space is not well understood yet. In the following we present this process in full details.

The Lagrangian of ZDG model is

$$L = -\sigma e.R(\omega) - e'.R(\omega') + \frac{m^2}{2}(\beta_1 e.e \times e' + \beta_2 e'.e' \times e), \quad (112)$$

where the primed fields represent second gravity fields and interaction terms appear by multipliers $\beta_1$ and $\beta_2$. Comparing with Eq. (35), the nonvanishing coefficient $g_{rs}$ and $f_{rst}$ are as follows

$$g_{e\omega} = -\sigma, \quad g_{e'\omega'} = -1,$$
$$f_{e\omega\omega} = -\sigma, \quad f_{e'\omega'\omega'} = -1,$$
$$f_{eee'} = m^2\beta_1, \quad f_{e'e'e} = m^2\beta_2. \quad (113)$$

The Lagrangian density corresponding to Lagrangian (112) is

$$\mathcal{L} = \frac{\sigma}{2}\epsilon^{ij}e_i.\dot{\omega}_j + \frac{\sigma}{2}\epsilon^{ij}\omega_i.\dot{e}_j + \frac{1}{2}\epsilon^{ij}e'_i.\dot{\omega}'_j + \frac{1}{2}\epsilon^{ij}\omega'_i.\dot{e}'_j$$
$$+ e_0.\phi_e + \omega_0.\phi_\omega + \acute{e}_0.\phi_{e'} + \acute{\omega}_0.\phi_{\omega'}, \quad (114)$$

where $\phi$'s are primary constraints as follows

$$\phi_\omega^a = -\sigma\epsilon^{ij}\partial_i e_j^a - \frac{\sigma}{2}\epsilon^{ij}\epsilon_{bc}^a e_i^b \omega_j^c,$$

$$\phi_{\omega'}^a = -\epsilon^{ij}\partial_i \acute{e}_j^a - \frac{1}{2}\epsilon^{ij}\epsilon_{bc}^a \acute{e}_i^b \acute{\omega}_j^c,$$

$$\phi_e^a = -\sigma\epsilon^{ij}\partial_i \omega_j^a - \frac{\sigma}{2}\epsilon^{ij}\epsilon_{bc}^a \omega_i^b \omega_j^c + m^2\frac{\beta_1}{2}\epsilon^{ij}\epsilon_{bc}^a e_i^b \acute{e}_j^c$$
$$+ m^2\frac{\beta_2}{2}\epsilon^{ij}\epsilon_{bc}^a \acute{e}_i^b \acute{e}_j^c,$$

$$\phi_{e'}^a = -\epsilon^{ij}\partial_i \acute{\omega}_j^a - \frac{1}{2}\epsilon^{ij}\epsilon_{bc}^a \acute{\omega}_i^b \acute{\omega}_j^c + m^2\frac{\beta_2}{2}\epsilon^{ij}\epsilon_{bc}^a \acute{e}_i^b e_j^c$$
$$+ m^2\frac{\beta_1}{2}\epsilon^{ij}\epsilon_{bc}^a e_i^b e_j^c. \quad (115)$$

The kinetic term in Lagrangian density (114) determines nonvanishing Poisson brackets as

$$\{e_i^a(\mathbf{x}), \omega_j^b(\mathbf{x}')\} = \{\acute{e}_i^a(\mathbf{x}), \acute{\omega}_j^b(\mathbf{x}')\} = \epsilon_{ij}\eta^{ab}\delta^2(\mathbf{x}-\mathbf{x}'). \quad (116)$$

In Appendix A we calculate Poisson brackets among primary constraints $\phi$. In contrast to TMG and NMG discussed above, here neither $\phi_\omega$ nor $\phi_{\omega'}$ commute with the other constraints. However, on the basis of physical grounds we expect that the combination $\phi_W^a = \phi_\omega^a + \phi_{\omega'}^a$ act as the generator of LLT's. Fortunately our guess is agreed directly by using the Poisson brackets given in appendix A. We can see that $\phi_W$ are first class. So, as before, the LLT's are unlimited gauge transformation through the whole phase space.

Using our experiences in TMG and NMG, let consider the generator of diffeomorphism, which as before [see Eqs. (68) and (95)] read

$$\psi_\mu = e_\mu^a \phi_e^a + \acute{e}_\mu^a \phi_{e'}^a + \omega_\mu^a \phi_\omega^a + \acute{\omega}_\mu^a \phi_{\omega'}^a. \quad (117)$$

We need to consider 6 more constraints together with $\phi_W$ and $\psi_\mu$ to form a new basis for the constraint surface [instead of the constraints of Eq. (115)]. A simple choice may be $\phi_\omega$ itself and say $\phi_k^a = \beta_1\phi_e^a - \beta_2\phi_{e'}^a$. The algebra of the constraints in the new basis of $\psi_\mu, \phi_k, \phi_W$, and $\phi_\omega$ is as follows

$$\{\phi_\omega^a, \phi_\omega^b\} = 0,$$
$$\{\psi_\mu, \phi_k^b\} \approx (\beta_1 e_\mu^b + \beta_2 \hat{e}_\mu^b) \epsilon^{ij}(\omega_i - \omega_i').(\beta_1 e_j + \beta_2 e_j') e_\mu^b,$$
$$\{\phi_k^a, \phi_k^b\} \approx \beta_1^2 \beta_2 \epsilon_c^{ab} \phi_e^c + \beta_2^2 \beta_1 \epsilon_c^{ab} \phi_{e'}^c,$$
$$\{\phi_k^a, \phi_\omega^b\} \approx \beta_1 \beta_2 \epsilon^{ij} e_i^b \hat{e}_j^a - \beta_1 \beta_2 \epsilon^{ij} e_i^a \hat{e}_j^b - \beta_1^2 \epsilon^{ij} e_i^a e_j^b$$
$$- \beta_2^2 \epsilon^{ij} \hat{e}_i^a \hat{e}_j^b,$$
$$\{\psi_\mu, \phi_\omega^b\} \approx (\beta_1 e_\mu^b + \beta_2 \hat{e}_\mu^b) \epsilon^{ij}(e_i.e_j'),$$
$$\{\psi_\mu, \psi_\nu\} \approx (e_\mu^a \hat{e}_\nu^a + e_\nu^a \hat{e}_\mu^a) \epsilon^{ij}(\omega_i - \omega_i').(\beta_1 e_j + \beta_2 e_j')$$
$$+ (\beta_1 e_\mu^b + \beta_2 \hat{e}_\mu^b)(\omega_\nu + \omega_\nu') \epsilon^{ij}(e_i.e_j'),$$
$$+ (e_\nu^b + \hat{e}_\nu^b)(\omega_\mu + \omega_\mu') \epsilon^{ij}(e_i.e_j'). \tag{118}$$

The requirement that $\psi_\mu$ be first class in the physical sector of the theory is fulfilled if we impose two new constraints $\Gamma$ and $\Omega$ as

$$\Gamma \equiv \epsilon^{ij}(e_i.e_j') \approx 0,$$
$$\Omega \equiv \epsilon^{ij}(\omega_i - \omega_i').(\beta_1 e_j + \beta_2 e_j') \approx 0. \tag{119}$$

It can be seen from Eqs. (118) that on the surface the constraints (including new constraints) the rank of the matrix of Poisson brackets is 4. This means that among the constraints $\phi_\omega$ and $\phi_k$ four ones have been remained second class on the surface of new constraints. On the other hand, the new constraints $\Gamma$ and $\Omega$, as in the case of NMG, do not commute with the other constraints. We may choose the remaining two constraints among $\phi_\omega$ and $\phi_k$ as conjugates of the new constraints. Hence, the set of constraints $\phi_\omega$, $\phi_k$, $\Gamma$ and $\Omega$ constitute a system of second class constraints. In this way the system possesses 8 second class and 6 first class constraints. The number of dynamical degrees of freedom from Eq. (21) read

$$24 - 6 \times 2 - 8 \times 1 = 4, \tag{120}$$

which is the desired result in agreement with Ref. [12].

Again it can be easily checked that by inserting the coefficients given in Eq. (114) for ZDG and by suitable choice of indices in Eq. (47), one may find the new constraints (119) in an other way. However, it worth emphasize again that the constraints (119) are not the natural consequence of the consistency procedure of the constraints, unless one employs additional assumption of the invertibility of the veilbeins.

## VII. CONCLUSIONS

In an ordinary constrained system a constraint is either primary which emerge due to singularity of the Lagrangian or is secondary which results from the Poisson brackets of the primary constraints with the canonical Hamiltonian [13,14,31]. It should be noted that every algebraic equation which is the result of mathematical manipulations of the primary constraints can not be considered as a secondary constrained. In fact, the way a constraint appears in the process of Hamiltonian analysis of a system is of great importance, for example in the role of that constraint in generating symmetries of the system.

In this paper we found a new feature of constrained system which was not recognized yet. It was well known in some older references of constrained systems [26] that the rank of the matrix of Poisson brackets may vary in different parts of the phase space. However, to the best of our knowledge, this point has not been considered as a new source of emerging constraints in a system.

We saw that by factorizing the determinant of the matrix of Poisson brackets of the second class constraints and indicating the "vanishable factors," one may find a new set of constraints. These constraints define some limited subspace of the phase space on which the physical properties, including symmetries of the system, may differ from elsewhere of phase space.

For CSL theories the dynamical properties of the system is not interesting in the ordinary constraint surface of the theory. In fact it gives a singular metric. However, we found that the rank of the matrix of Poisson bracket of second class constraints with non vanishing ordinary determinant changes on the surface of some new constraints. Fortunately on this subspace, first, the vielbeins are invertible and give a nonsingular metric and second, the diffeomorphism gauge symmetry finds its generators as new first class constraints.

We also observed that the linearized version of these models live on the physical sector of the system and naturally have enough number of degrees of freedom, i.e., the same as ordinary system in the physical sector.

We worked with zero cosmological constant. Treating with a nonzero cosmological constant is achieved, in vielbein framework, by adding the term $(\frac{\Lambda}{6}\varepsilon^{abc} e_a e_b e_c)$ to the Lagrangian of TMG and NMG and two similar copies to ZDG. Since this term is not kinetic, the fundamental Poisson brackets among the fields do not change. The only change is appearance of the additional term $\frac{\Lambda}{2} \epsilon^{ij} \epsilon_{abc} e_i^b e_j^c$ in the constraint $\phi_e^a$ in all three models considered (plus a similar term in $\phi_{e'}^a$ in ZDG). Direct calculation shows that the new terms do not make any change in the constraint structure of the original models. Although the dynamics of the dynamical variables (which have been remained in the reduced phase space) would change considerably, similarity in the constraint structure leads to the same number of degrees of freedom. Hence, we conclude that our main results do not change for the original models. However, for the linearized model one should expand the fields around the AdS or de Sitter solutions, depending on the sign of the cosmological constant. The dynamical structure of the linearized models in each case is well known and the number of degrees of freedom agree with the above results for zero cosmological constant.

## ACKNOWLEDGMENTS

The authors would like to thank Hamid R. Afshar, M. Alishahiha, and A. Naseh for helpful discussions and comments. They also thank M. Khodaei, R. Shiri, and M. Sadegh for their efforts in the earlier steps of this work.

## APPENDIX A: CONSISTENCY OF CONSTRAINTS

In this Appendix we illustrate some details that appear during the process of consistency of constraints. As we said in the text, since the total Hamiltonian is a linear combination of the constraints, consistency of constraint $\phi_r^a(\mathbf{x})$ leads to Eq. (43) as

$$\int d^2\mathbf{x}' \{\phi_r^a(\mathbf{x}), \phi_s^b(\mathbf{x}')\} a_0^{sb}(\mathbf{x}') \approx 0. \quad (A1)$$

Using Eq. (39) the general form of the primary constraints is as follows

$$\phi_r^a(\mathbf{x}) = \varepsilon^{ij} g_{rm} \partial_i a_j^{ma}(\mathbf{x}) + \frac{1}{2} \varepsilon^{ij} \varepsilon^a_{cd} f_{rst} a_i^{sc}(\mathbf{x}) a_j^{td}(\mathbf{x}). \quad (A2)$$

It is obvious from Eq. (A2) that each constraint includes two types of terms, the terms with spatial derivative of the fields and the terms that are quadratic with respect to the fields. Therefore, in the process of obtaining the Poisson bracket between two constraints three kinds of terms emerge, i.e., {derivative, derivative}, {quadratic, quadratic}, and {derivative quadratic}. The second kind of terms are again quadratic terms (times delta function which disappear under integration over one set of spatial coordinates). For the other two kinds we should care about the derivatives of delta functions. The first kind of terms vanish due to the symmetries as follows

$$\begin{aligned}
&\{\varepsilon^{ij} g_{rs} \partial_i a_j^{sa}(\mathbf{x}), \varepsilon^{kl} g_{tu} \partial'_k a_l^{ub}(\mathbf{x}')\} \\
&= \varepsilon^{ij} \varepsilon^{kl} g_{rs} g_{tu} \partial_i \partial'_k \{a_j^{sa}(\mathbf{x}), a_l^{tb}(\mathbf{x}')\} \\
&= \varepsilon^{ij} \varepsilon^{kl} g_{rs} g_{tu} \varepsilon_{jl} \eta^{ab} g^{st} \partial_i \partial'_k \delta^2(\mathbf{x} - \mathbf{x}') \\
&= 2 g_{rs} g_{tu} g^{st} \eta^{ab} \varepsilon^{ki} \partial_i \partial_k \delta^2(\mathbf{x} - \mathbf{x}') = 0. \quad (A3)
\end{aligned}$$

The most important part of calculation corresponds to the third kind of terms. Let us consider, for example, a term like this (appearing in $\{\phi_\omega^1(\mathbf{x}), H_T\}$ of the TMG)

$$\int d^2\mathbf{x}' \omega_0^2(\mathbf{x}') \{\partial_1 e_2^1(\mathbf{x}), -e_2^3(\mathbf{x}') h_1^1(\mathbf{x}')\}$$
$$= \int d^2\mathbf{x}' \omega_0^2(\mathbf{x}') e_2^3(\mathbf{x}') \partial_1 \delta^2(\mathbf{x} - \mathbf{x}'). \quad (A4)$$

The interesting point is that we can find a partner term in the same Poisson bracket which gives similar result in which $\partial_i \delta^2(\mathbf{x} - \mathbf{x}')$ is replaced by $\partial'_i \delta^2(\mathbf{x} - \mathbf{x}')$ and the argument of the dynamical field changes. Fortunately this is the case for all Chern-simons-like models considered in this paper. For example in $\{\phi_\omega^1(\mathbf{x}), H_T\}$ we have also the following term

$$\int d^2\mathbf{x}' \omega_0^2(\mathbf{x}') \{e_2^3(\mathbf{x}) h_1^2(\mathbf{x}), \partial'_1 e_2^1(\mathbf{x}')\}$$
$$= \int d^2\mathbf{x}' \omega_0^2(\mathbf{x}') e_2^3(\mathbf{x}') \partial'_1 \delta^2(\mathbf{x} - \mathbf{x}'). \quad (A5)$$

Summing the above terms, and putting the result in the expression of $\{\phi_\omega^1(\mathbf{x}), H_T\}$, which contains multiplication with undetermined function $\omega_0^2$ and finally integration over $\mathbf{x}'$, we find

$$\{\phi_\omega^1(\mathbf{x}), H\} = \cdots + \int \omega_0^2(\mathbf{x}') \partial'_1 \delta^2(\mathbf{x} - \mathbf{x}') (e_2^3(\mathbf{x}) - e_2^3(\mathbf{x}')) d^2\mathbf{x}', \quad (A6)$$

where dots represent other terms in the expression. Integrating by parts, the corresponding term in $\{\phi_\omega, H\}$ reads

$$\int d^2\mathbf{x}' \omega_0^2(\mathbf{x}') \partial'_1 \delta^2(\mathbf{x} - \mathbf{x}') (e_2^3(\mathbf{x}) - e_2^3(\mathbf{x}'))$$
$$= -\int d^2\mathbf{x}' \omega_0^2(\mathbf{x}') \delta^2(\mathbf{x} - \mathbf{x}') \partial'_1 (e_2^3(\mathbf{x}) - e_2^3(\mathbf{x}'))$$
$$- \int d^2\mathbf{x}' \partial'_1 \omega_0^2(\mathbf{x}') \delta^2(\mathbf{x} - \mathbf{x}') (e_2^3(\mathbf{x}) - e_2^3(\mathbf{x}')). \quad (A7)$$

The last term in Eq. (A7) vanishes due to the delta function. We have finally

$$\int d^2\mathbf{x}' \omega_0^2(\mathbf{x}') \partial'_1 \delta^2(\mathbf{x} - \mathbf{x}') (e_2^3(\mathbf{x}) - e_2^3(\mathbf{x}'))$$
$$= \omega_0^2(\mathbf{x}) \partial_1 e_2^3(\mathbf{x}). \quad (A8)$$

In this way we can write the following relation:

$$\{\text{derivative, quadratic}\} + \{\text{quadratic, derivative}\}$$
$$= \text{derivative} \times \delta^2(\mathbf{x} - \mathbf{x}'). \quad (A9)$$

Hence, the Poisson brackets among the constraints give expressions which have the same structure as the constraints, i.e., derivative + quadratic.

Direct calculation using the fundamental Poisson brackets (53) and inserting the particular values (49) of $g_{rs}$ and $f_{rst}$ for the case of TMG gives the following result

$$\{\phi_\omega^a(x_\mu), \phi_\omega^b(x'_\mu)\} = \epsilon_c^{ab} \phi_\omega^c(x_\mu),$$

$$\{\phi_\omega^a(x_\mu), \phi_h^b(x'_\mu)\} = \epsilon_c^{ab} \phi_h^c(x_\mu),$$

$$\{\phi_\omega^a(x_\mu), \phi_e^b(x'_\mu)\} = \epsilon_c^{ab} \phi_e^c(x_\mu),$$

$$\{\phi_e^a(x_\mu), \phi_h^b(x'_\mu)\} = -\epsilon_c^{ab} \epsilon^{ij} \partial_i \omega_j^c(x_\mu) - \frac{\mu}{2} \epsilon^{ij} e_i^c h_j^c(x_\mu) \delta^{ab}$$
$$- \frac{\mu}{2} \epsilon^{ij} e_i^a(x_\mu) h_j^b(x_\mu) - \frac{1}{2} \epsilon^{ij} \omega_i^a(x_\mu) \omega_j^b(x_\mu),$$

$$\{\phi_e^a(x_\mu), \phi_e^b(x'_\mu)\} = -\frac{\mu}{2} \epsilon^{ij} h_i^a(x_\mu) h_j^b(x_\mu),$$

$$\{\phi_h^a(x_\mu), \phi_h^b(x'_\mu)\} = -\frac{\mu}{2} \epsilon^{ij} e_i^a(x_\mu) e_j^b(x_\mu). \quad \text{(A10)}$$

For the new constraint $\Gamma \equiv e.h = 0$ we have

$$\{\Gamma(x_\mu), \phi_\omega^a(x'_\mu)\} = 0,$$

$$\{\Gamma(x_\mu), \phi_h^a(x'_\mu)\} = \epsilon^{ij} \partial_i e_j^a(x_\mu) + \frac{1}{2} \epsilon_{bc}^a \epsilon^{ij} e_i^b e_j^c$$
$$+ \frac{1}{2} \epsilon_{bc}^a \epsilon^{ij} e_i^b \omega_j^c,$$

$$\{\Gamma(x_\mu), \phi_e^a(x'_\mu)\} = \epsilon^{ij} \partial_i h_j^a(x_\mu) + \frac{1}{2} \epsilon_{bc}^a \epsilon^{ij} e_i^b h_j^c$$
$$+ \frac{1}{2} \epsilon_{bc}^a \epsilon^{ij} h_i^b \omega_j^c. \quad \text{(A11)}$$

A multiplicative factor of $\delta^2(\mathbf{x} - \mathbf{x}')$ should be understood in the right-hand side of all above and below Poisson brackets.

For NMG with fundamental Poisson brackets in Eq. (88) we find

$$\{\phi_\omega^a(x_\mu), \phi_\omega^b(x'_\mu)\} = \epsilon_c^{ab} \phi_\omega^c(x_\mu),$$

$$\{\phi_\omega^a(x_\mu), \phi_h^b(x'_\mu)\} = \epsilon_c^{ab} \phi_h^c(x_\mu),$$

$$\{\phi_\omega^a(x_\mu), \phi_e^b(x'_\mu)\} = \epsilon_c^{ab} \phi_e^c(x_\mu),$$

$$\{\phi_\omega^a(x_\mu), \phi_f^b(x'_\mu)\} = \epsilon_c^{ab} \phi_f^c(x_\mu),$$

$$\{\phi_e^a(x_\mu), \phi_h^b(x'_\mu)\} = \epsilon_c^{ab} \epsilon^{ij} \partial_i \omega_j^c(x_\mu) - \frac{\epsilon^{ij}}{2} f_i^c(x_\mu) e_j^c(x_\mu) \delta^{ab}$$
$$- \frac{\epsilon^{ij}}{2} \omega_i^a(x_\mu) \omega_j^b(x_\mu) - \frac{\epsilon^{ij}}{2} e_i^a(x_\mu) f_j^b(x_\mu),$$

$$\{\psi_e^a(x_\mu), \psi_f^b(x'_\mu)\} = \epsilon_c^{ab} \epsilon^{ij} \partial_i f_j^c(x_\mu) + \frac{\epsilon^{ij}}{2} e_i^c(x_\mu) h_j^c(x_\mu) \delta^{ab}$$
$$+ \frac{\epsilon^{ij}}{2m^2} f_i^a(x_\mu) \omega_j^b(x_\mu) - \frac{\epsilon^{ij}}{2m^2} f_i^b(x_\mu) \omega_j^a(x_\mu)$$
$$+ \frac{1}{2} \epsilon^{ij} e_i^b(x_\mu) h_j^a(x_\mu),$$

$$\{\phi_e^a(x_\mu), \phi_e^b(x'_\mu)\} = \frac{1}{2} \epsilon^{ij} f_i^a(x_\mu) h_j^b(x_\mu) - \frac{1}{2} \epsilon^{ij} f_i^b(x_\mu) h_j^a(x_\mu),$$

$$\{\phi_h^a(x_\mu), \phi_f^b(x'_\mu)\} = \frac{1}{2} \epsilon^{ij} e_i^a(x_\mu) e_j^b(x_\mu),$$

$$\{\phi_h^a(x_\mu), \phi_h^b(x'_\mu)\} = 0,$$

$$\{\phi_f^a(x_\mu), \phi_f^b(x'_\mu)\} = \epsilon_c^{ab} \phi_h^c(x_\mu). \quad \text{(A12)}$$

Finally, Poisson brackets among primary constraints of ZDG are obtained as

$$\{\phi_e^a(x_\mu), \phi_e^b(x'_\mu)\} \approx m^2 \beta_1 \acute{e}_i^a(x_\mu) \omega_j^b(x_\mu) - m^2 \beta_1 \acute{e}_i^b(x_\mu) \omega_j^a(x_\mu) + m^2 \beta_1 \acute{e}_i^a(x_\mu) \acute{\omega}_j^b(x_\mu) - m^2 \beta_1 \acute{e}_i^b(x_\mu) \acute{\omega}_j^a(x_\mu),$$

$$\{\phi_e^a(x_\mu), \phi_\omega^b(x'_\mu)\} \approx m^2 \beta_1 \frac{\epsilon^{ij}}{2} e_i^c(x_\mu) \acute{e}_j^c(x_\mu) \delta^{ab} + m^2 \beta_1 \frac{\epsilon^{ij}}{2} \acute{e}_i^a(x_\mu) \acute{e}_j^b(x_\mu) - m^2 \beta_1 \frac{\epsilon^{ij}}{2} e_i^b(x_\mu) \acute{e}_j^a(x_\mu),$$

$$\{\phi_e^a(x_\mu), \phi_{e'}^b(x'_\mu)\} \approx m^2 \beta_1 \frac{\epsilon^{ij}}{2} e_i^c(x_\mu) \omega_j^c(x_\mu) \delta^{ab} - m^2 \beta_1 \frac{\epsilon^{ij}}{2} e_i^b(x_\mu) \omega_j^a(x_\mu) + m^2 \beta_1 \frac{\epsilon^{ij}}{2} e_i^c(x_\mu) \acute{\omega}_j^c(x_\mu) \delta^{ab}$$
$$+ m^2 \beta_1 \frac{\epsilon^{ij}}{2} e_i^b(x_\mu) \acute{\omega}_j^a(x_\mu) + m^2 \beta_2 \frac{\epsilon^{ij}}{2} \acute{e}_i^c(x_\mu) \omega_j^c(x_\mu) \delta^{ab} - m^2 \beta_2 \frac{\epsilon^{ij}}{2} \acute{e}_i^a(x_\mu) \omega_j^b(x_\mu)$$
$$+ m^2 \beta_2 \frac{\epsilon^{ij}}{2} \acute{e}_i^c(x_\mu) \acute{\omega}_j^c(x_\mu) \delta^{ab} - m^2 \beta_2 \frac{\epsilon^{ij}}{2} \acute{e}_i^a(x_\mu) \acute{\omega}_j^b(x_\mu),$$

$$\{\phi_e^a(x_\mu), \phi_{\omega'}^b(x'_\mu)\} \approx -\frac{\epsilon^{ij}}{2} e_i^c(x_\mu) \acute{e}_j^c(x_\mu) \delta^{ab} + m^2 \beta_1 \frac{\epsilon^{ij}}{2} e_i^b(x_\mu) \acute{e}_j^a(x_\mu) + m^2 \beta_2 \frac{\epsilon^{ij}}{2} \acute{e}_i^a(x_\mu) \acute{e}_j^b(x_\mu),$$

$$\{\phi_\omega^a(x_\mu), \phi_\omega^b(x'_\mu)\} = 0,$$

$$\{\phi_\omega^a(x_\mu), \phi_{e'}^b(x'_\mu)\} \approx -m^2 \beta_2 \frac{\epsilon^{ij}}{2} e_i^c(x_\mu) \acute{e}_j^c(x_\mu) \delta^{ab} + m^2 \beta_2 \frac{\epsilon^{ij}}{2} e_i^b(x_\mu) \acute{e}_j^a(x_\mu) + m^2 \beta_1 \frac{\epsilon^{ij}}{2} e_i^a(x_\mu) e_j^b(x_\mu),$$

$$\{\phi_\omega^a(x_\mu), \phi_{\omega'}^b(x'_\mu)\} = 0,$$

$$\{\phi_{e'}^a(x_\mu), \phi_{e'}^b(x'_\mu)\} \approx m^2 \beta_2 e_i^a(x_\mu) \acute{\omega}_j^b(x_\mu) - m^2 \beta_2 e_i^b(x_\mu) \acute{\omega}_j^a(x_\mu) + m^2 \beta_2 e_i^a(x_\mu) \omega_j^b(x_\mu) - m^2 \beta_2 e_i^b(x_\mu) \omega_j^a(x_\mu)$$

$$\{\phi_{e'}^a(x_\mu), \phi_{\omega'}^b(x'_\mu)\} \approx m^2 \beta_2 \frac{\epsilon^{ij}}{2} \acute{e}_i^c(x_\mu) e_j^c(x_\mu) \delta^{ab} + m^2 \beta_2 \frac{\epsilon^{ij}}{2} e_i^a(x_\mu) e_j^b(x_\mu) - m^2 \beta_2 \frac{\epsilon^{ij}}{2} \acute{e}_i^b(x_\mu) e_j^a(x_\mu),$$

$$\{\phi_{\omega'}^a(x_\mu), \phi_{\omega'}^b(x'_\mu)\} = 0. \quad \text{(A13)}$$

## APPENDIX B: SYMMETRIES AND GENERATORS

In the process of analyzing the CSL theories in form formulation, we expect to find two gauge transformation related to two symmetries of the theory, i.e., Local lorentz transformation which exist because we have freedom to chose local Lorentz frame and diffeomorphism which is the basic symmetry of covariant system and exist in the metric formulation too. In this Appendix we show that the generators of these symmetries are the specified first class constraints (indicated in the text), which emerge in the canonical structure of the system.

Let begin with LLT symmetry. It is well known [32] that under LLT infinitesimal gauge transformation basic fields $e$ and $\omega$ vary as follows

$$\delta e_\mu^a = -\alpha_b^a e_\mu^b,$$
$$\delta \omega_\mu^a = -\alpha_b^a e_\mu^b + \frac{1}{2} \epsilon^{abc} \partial_\mu \alpha_{bc}, \quad (B1)$$

and other auxiliary one form vary like

$$\delta A_\mu^a = -\alpha_b^a A_\mu^b, \quad (B2)$$

where $\alpha_b^a$ is antisymmetric infinitesimal lorentz parameter. In the models we studied, first class constraints which generate unlimited gauge transformation are generator of LLT symmetry. This includes $\phi_\omega$ for TMG and NMG and $\phi_W$ for ZDG. Assume the following gauge generating functional for TMG and NMG

$$\Phi_\omega = \int \varepsilon_{akl} \alpha^{kl} \phi_\omega^a d^2x, \quad (B3)$$

and the following for ZDG

$$\Phi_W = \int \varepsilon_{akl} \alpha^{kl} (\phi_\omega^a + \phi_{\omega'}^a) d^2x. \quad (B4)$$

Using the fundamental Poisson brackets (53) and (89) given in the text and the explicit form of the constraints $\phi_\omega$ in Eqs. (52) and (87) we find directly the following relations for NMG

$$\{\Phi_\omega, e_a^\mu\} = -\alpha_b^a e_\mu^b,$$
$$\{\Phi_\omega, h_a^\mu\} = -\alpha_b^a h_\mu^b,$$
$$\{\Phi_\omega, f_a^\mu\} = -\alpha_b^a f_\mu^b,$$
$$\{\Phi_\omega, \omega_a^\mu\} = -\alpha_b^a \omega_\mu^b + \frac{1}{2} \epsilon^{abc} \partial_\mu \alpha_{bc} \quad (B5)$$

and similar relations for TMG, except the third one. Similar treatments for ZDG by using the fundamental Poisson brackets (116) and explicit form of the constraints $\phi_\omega$ and $\phi_{\omega'}$ from (115) gives

$$\{\Phi_W, e_a^\mu\} = -\alpha_b^a e_\mu^b,$$
$$\{\Phi_W, \omega_a^\mu\} = -\alpha_b^a \omega_\mu^b + \frac{1}{2} \epsilon^{abc} \partial_\mu \alpha_{bc},$$
$$\{\Phi_W, e_a'^\mu\} = -\alpha_b^a e_\mu'^b,$$
$$\{\Phi_W, \omega_a'^\mu\} = -\alpha_b^a \omega_\mu'^b + \frac{1}{2} \epsilon^{abc} \partial_\mu \alpha_{bc}. \quad (B6)$$

The right-hand side of Eqs. (B5) and (B6) are exactly in the expected form of Eqs. (B1) and (B2). This shows that $\Phi_\omega$ and $\Phi_W$ given in Eqs. (B3) and (B4) are generators of LLT in TMG, NMG and ZDG respectively.

About diffeomorphism symmetry, it is well known that variation of any one form $A_\mu$ under infinitesimal gauge transformation is

$$\delta A_\mu = \xi^\nu \partial_\nu A_\mu + A_\nu \partial_\mu \xi^\nu = \xi^\nu (\partial_\nu A_\mu - \partial_\mu A_\nu) + \partial_\mu (A_\nu \xi^\nu), \quad (B7)$$

where the vector $\xi^\nu$ indicates the infinitesimal diffeomorphism variation of coordinates as $x^\mu \to x^\mu - \xi^\mu$. Let us define the generating functional $\Psi$ as

$$\Psi = \int \xi^\mu \psi_\mu d^2x, \quad (B8)$$

where generators $\psi_\mu$ are given in Eqs. (68), (95), and (117) for TMG, NMG, and ZDG respectively.

Using the fundamental Poisson brackets it can directly verified that

$$\{\Psi, e_a^\mu\} = \xi^\nu \partial_\nu e_\mu^a + e_\nu^a \partial_\mu \xi^\nu,$$
$$\{\Psi, h_a^\mu\} = \xi^\nu \partial_\nu h_\mu^a + h_\nu^a \partial_\mu \xi^\nu,$$
$$\{\Psi, f_a^\mu\} = \xi^\nu \partial_\nu f_\mu^a + f_\nu^a \partial_\mu \xi^\nu,$$
$$\{\Psi, \omega_a^\mu\} = \xi^\nu \partial_\nu \omega_\mu^a + \omega_\nu^a \partial_\mu \xi^\nu. \quad (B9)$$

for TMG (except the third relation) and NMG. While for ZDG we find

$$\{\Psi, e_a^\mu\} = \xi^\nu \partial_\nu e_\mu^a + e_\nu^a \partial_\mu \xi^\nu,$$
$$\{\Psi, \omega_a^\mu\} = \xi^\nu \partial_\nu \omega_\mu^a + \omega_\nu^a \partial_\mu \xi^\nu,$$
$$\{\Psi, \acute{e}_a^\mu\} = \xi^\nu \partial_\nu \acute{e}_\mu^a + \acute{e}_\nu^a \partial_\mu \xi^\nu,$$
$$\{\Psi, \acute{\omega}_a^\mu\} = \xi^\nu \partial_\nu \acute{\omega}_\mu^a + \acute{\omega}_\nu^a \partial_\mu \xi^\nu. \quad (B10)$$

Comparing to Eq. (B7), Eqs. (B9) and (B10) indicate that $\psi_\mu$ in all cases considered are generators of diffeomorphism.

HAMILTONIAN STRUCTURE OF THREE-DIMENSIONAL … PHYS. REV. D **97,** 024022 (2018)